\documentclass[a4paper]{tMOP2e}

\usepackage{amsmath}
\usepackage[OT1]{fontenc}

\newcommand{\rmi}{\mathrm{i}}
\newcommand{\rmd}{\mathrm{d}}
\newcommand{\Mod}{\mathrm{mod}\,}

\makeatletter
\newcommand{\@doi}{\today}
\makeatother

\begin{document}

\markboth{J. B. G{\"o}tte et al.}{Quantum Formulation of Fractional Orbital Angular Momentum}

%%%%%%%%%%%%%%%%% title page information %%%%%%%%%%%%%%%%%%
\title{Quantum Formulation of Fractional Orbital Angular Momentum}

\author{\uppercase{J. B. G{\"o}tte$^\dagger$, S. Franke-Arnold$^\ddagger$, R. Zambrini$^\S$ and Stephen M. Barnett$^\dagger$} \\
$^\dagger$SUPA, University of Strathclyde, Department of Physics, Glasgow G4 0NG, United Kingdom \\
$^\ddagger$SUPA, University of Glasgow, Department of Physics and Astronomy, Glasgow G12 8QQ, United Kingdom \\
$^\S$IMEDEA, CSIC-UIB, Campus de la Universitat de las Illes Baleares, Palma de Mallorca, E-07122, Spain}

\maketitle

%\homepage{http://cnqo.phys.strath.ac.uk/~joerg} %% author's URL, if desired

%%%%%%%%%%%%%%%%%%% abstract and OCIS codes %%%%%%%%%%%%%%%%

\begin{abstract}
The quantum theory of rotation angles (S. M. Barnett and D. T. Pegg, Phys. Rev. A, \textbf{41}, 3427-3425 (1990)) is generalised to non-integer values of the orbital angular momentum. This requires the introduction of an additional parameter, the orientation of a phase discontinuity associated with fractional values of the orbital angular momentum. We apply our formalism to the propagation of light modes with fractional orbital angular momentum in the paraxial and non-paraxial regime.
\end{abstract} 

%%%%%%%%%%%%%%%%%%%%%%% References %%%%%%%%%%%%%%%%%%%%%%%%%

%%%%%%%%%%%%%%%%%%%%%%%%%%  body  %%%%%%%%%%%%%%%%%%%%%%%%%%
\section{Introduction}
The orbital angular momentum of light beams is a consequence of their azimuthal phase structure \cite{allen+:pra45:1992,allen+:iop:2003}. Light beams with a phase factor  $\exp(\rmi m \phi)$, where $m$ is an integer and $\phi$ is the azimuthal angle, carry orbital angular momentum (OAM) of $m\hbar$ per photon along the beam axis \cite{allen+:pra45:1992}. These light beams can be generated in the laboratory by optical devices which manipulate the phase of the beam, such as spiral phase plates or holograms. In cases where such a device generates a light beam with an integer value of $m$, the resulting phase structure
has the form of $|m|$ intertwined helices of equal phase. For integer values of $m$ the chosen height of the phase step generated by the optical device is equal to the mean value of the OAM in the resulting beam. Recently, spiral phase steps with fractional step height \cite{oemrawsingh+:joa6:2004} as well as special holograms \cite{leach+:njp6:2004} have been used to generate light beams with fractional OAM. 
Here, the generating optical device imposes a phase change of $\exp(\rmi M \phi)$ and $M$ is not restricted to integer values. The phase structure of such beams shows a far more complex pattern; a series of optical vortices with alternating charge is created in a dark line along the direction of the phase discontinuity imprinted by the optical device \cite{berry:joa6:2004, leach+:njp6:2004}. 
In order to obtain the mean value of the orbital angular momentum of these beams one has to average over the vortex pattern. 
Interestingly, this mean value coincides with the phase-step only for integer and half-integer values. Our aim is to exploit the well-known formal connections between optics and quantum theory to represent beams with fractional OAM as quantum states. We anticipate that this will be particularly useful, when applied to single photons and related quantum phenomena.

Until now, the theoretical description of light modes with fractional OAM has been based on the generating optical device. For integer OAM values, however, a fundamental theoretical description exists which provides the only known way to treat the angle itself as quantum mechanical Hermitian operator \cite{barnettpegg:pra41:1990}. This description provides the underlying theory for a secure quantum communication system \cite{gibson+:oe12:2004} and it gives a form of the uncertainty relation for angle and angular momentum which has been confirmed in an optical experiment \cite{frankearnold+:njp6:2004}. In this paper we generalise this theory to fractional values of $M$ thereby creating a quantum mechanical description of fractional OAM. Such a rigorous formulation is of particular interest as the use of half-integer spiral phase plates has been reported recently in actual and proposed experimental schemes to study high-dimensional entanglement \cite{oemrawsingh+:prl92:2004,aiello+:pra72:2005,oemrawsingh+:prl95:2005}. These studies have shown that fractional OAM states are characterised not only by the height of the phase step but also by the orientation of the phase dislocation $\alpha$. For half odd-integer values of $M$, that is for $M \Mod 1 = 1/2$, states with the same $M$ but with a $\pi$ difference in $\alpha$ are orthogonal.  In light of recent applications of integer OAM in quantum key distribution \cite{spedalieri:oc260:2006} and the conversion of spin to orbital angular momentum in an optical medium \cite{marrucci+:prl96:2006} a rigorous formulation is important for possible applications of fractional OAM to quantum communication.
 
\section{Fractional orbital angular momentum}
The component of the OAM in the propagation direction $L_z$ and the azimuthal rotation angle form a pair of conjugate variables \cite{pegg+:jmo37:1990}. Unlike linear position and momentum, which are both defined on an unbound and continuous state space, the state spaces for OAM and the rotation angle are different in nature; whereas the OAM eigenstates form a discrete set of states $\{|m\rangle\}_{m \in \mathbb{Z}}$ with $m$ taking on all integer values, eigenstates of the angle operator are restricted to a $2\pi$ radian interval, as it is physically impossible to distinguish between rotation angles differing by $2\pi$ radians. The properties of the angle operator are rigorously derived in an arbitrarily large, yet finite state space of $2L + 1$ dimensions \cite{barnettpegg:pra41:1990}. This space is spanned by the angular momentum states $| m \rangle$ with $m$ ranging from $-L,-L+1,\dots,L$. Accordingly, the $2\pi$ radian interval $[\theta_0, \theta_0+2\pi)$ 
is spanned by $2L+1$ orthogonal angle states $| \theta_n \rangle$ with $\theta_n = \theta_0 + 2\pi n/(2L+1)$. Here, $\theta_0$ determines the starting point of the interval and with it a particular angle operator $\hat{\phi}_\theta$. Only after physical results have been calculated within this state space is $L$ allowed to tend to infinity, which recovers the result of an infinite but countable number of basis states for the OAM and a dense set of angle states within a $2\pi$
radian interval. We conduct the analysis of fractional OAM in the finite dimensional space. After we have calculated expectation values or probabilities we can give final results in the limit of $L \to \infty$.

\begin{figure}
\begin{center}
\includegraphics[width=0.7\textwidth]{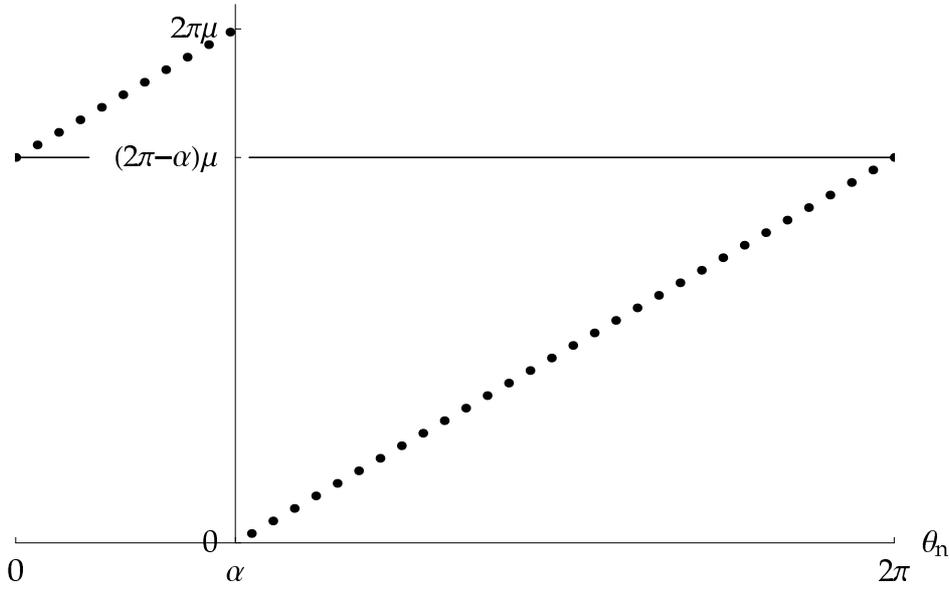}
\end{center}
\caption{\label{fig:phasejump}Plot of the phase discontinuity in the fractional OAM states $|M(\alpha)\rangle$. The relevant phase property is given by the factor $\exp[\rmi \mu( \theta_n +2\pi f_\alpha(\theta_n) - \alpha)]$. For the plot we have set $\theta_0 = 0$. If $\theta_0 \neq 0$ there is an additonal global phase factor $\exp(\rmi \mu \theta_0)$.}
\end{figure}

A quantum state with fractional OAM is denoted by $| M \rangle$, where $M = m  + \mu$ and $m$ is the integer part and $\mu \in [0,1)$ is the fractional part. We choose `fractional' rather than `non-integer'  to follow the terminology in earlier work, but $\mu$ can take on every real number $\in [0,1)$ not just rational numbers. This distinction, however, is not relevant for practical purposes. The state $| M \rangle$ is decomposed in angle states according to 
\begin{equation}
| M \rangle = \left( 2L + 1 \right)^{-1/2} \sum_{n = 0}^{2L} \exp(\rmi M \theta_n) | \theta_n \rangle = 
\left( 2L + 1 \right)^{-1/2} \sum_{n = 0}^{2L} \exp(\rmi m \theta_n) \exp(\rmi \mu \theta_n) | \theta_n \rangle.
\end{equation} 
In general $\exp(\rmi \mu \theta_n)$ will be a multivalued function but a common way to render a multi-valued function single-valued is the introduction of a branch cut \cite{conway:sv:1978}, which restricts the range of the function. Usually, the branch cut is chosen to be along the negative real axis in the complex plane, such that the function has no discontinuities. Here, we choose the position of the branch cut $\alpha$, that is the position of the discontinuity in the function $\exp(\rmi \mu \theta_n)$, independently of $\theta_0$, the starting angle of the $2\pi$ radian
interval. To achieve this we introduce an integer-valued function $f_\alpha(\theta_n)$ which takes on the values $1$ and $0$, depending on whether $\theta_n$ is smaller or greater than $\alpha$:
\begin{equation}
\label{eq:ffunction}
f_\alpha (\theta_n) : \{ \theta_n \}_{n = 0,\dots,2L} \longrightarrow \{ 0, 1 \}, 
\quad \theta_n \longmapsto 
\left\{ \begin{array}{lc} 1 & \theta_0 < \theta_n < \theta_0 + \alpha, \\ 0 & \theta_0 + \alpha < \theta_n < \theta_0 + 2\pi. 
\end{array} \right. 
\end{equation}
It is important to note that $\alpha$ is bounded by $0 \leq \alpha < 2\pi$, so that the orientation 
of the discontinuity is always understood as measured from $\theta_0$.
We should stress that according to the mathematical definition we do not create a branch of the multivalued function, as we explicitly include the discontinuity. With this construction the fractional state $| M \rangle$ can be written as
\begin{equation}
\label{eq:fracstatedef}
| M (\alpha) \rangle = \frac{\exp(-\rmi \mu \alpha)}{\sqrt{2L + 1}} \sum_{n = 0}^{2L} \exp(\rmi M \theta_n) 
\exp[\rmi 2\pi \mu f_\alpha(\theta_n)] | \theta_n \rangle.
\end{equation}
The fractional OAM state $|M (\alpha)\rangle$ clearly depends on the orientation of the discontinuity $\alpha$ (see figure \ref{fig:phasejump}), but the angle probability distribution is flat, $P(\theta_n) = \langle M(\alpha) | M(\alpha) \rangle = (2L+1)^{-1}$, as it is for the integer OAM states.  For the appropriate choice of input state, phase shift and orientation this is equivalent to the action of a spiral phase plate operator acting on an integer OAM state \cite{oemrawsingh+:prl92:2004}. It is formally possible to introduce a unitary operator $\hat{U}_m(\beta)$ which rotates the position of the discontinuity by writing $\hat{U}_m(\beta) | M(\alpha) \rangle = | M (\alpha \oplus \beta) \rangle$, where $M=m=\mu$ such that the index of the operator is matched by the integer part of $M$. We have introduced $\alpha \oplus \beta := (\alpha + \beta) \Mod 2\pi$ as short hand notation for the modulo $2\pi$ addition. The action of the operator $\hat{U}_m(\beta)$ on a general state $| M'(\alpha) \rangle$
results, not only in a rotation of the orientation, but also in a phase shift:
\begin{equation}
\hat{U}_m(\beta) | M' (\alpha) \rangle = \exp[\rmi(m-m')\beta] | M'(\alpha \oplus \beta) \rangle.
\end{equation}
Further properties of this operator are described in Appendix \ref{app:unitaryrotation}.

\subsection{Overlap of fractional OAM states}
\label{sec:overlap}

\begin{figure}
\begin{center}
 \includegraphics[width=0.7\textwidth]{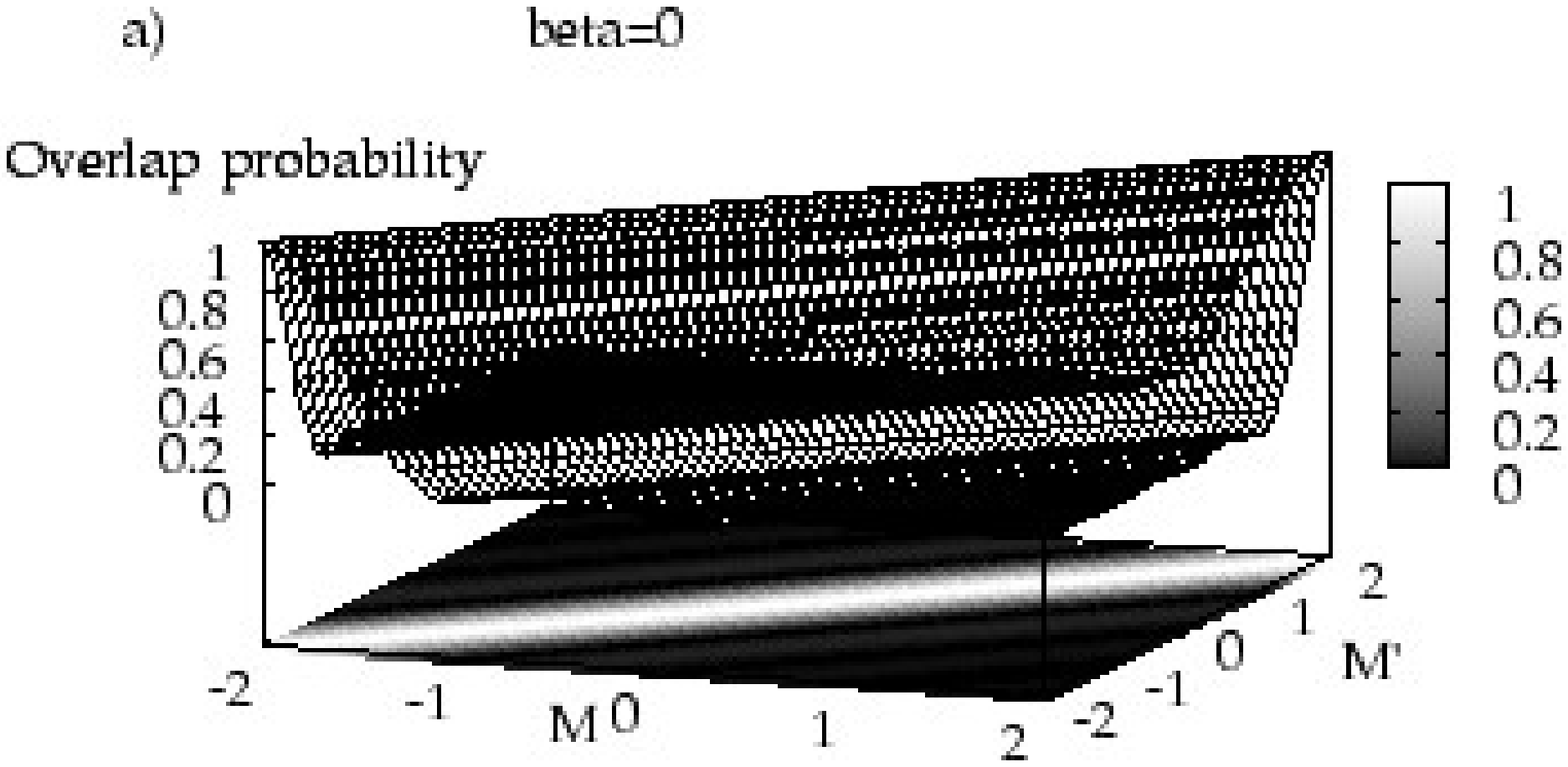}
 \includegraphics[width=0.7\textwidth]{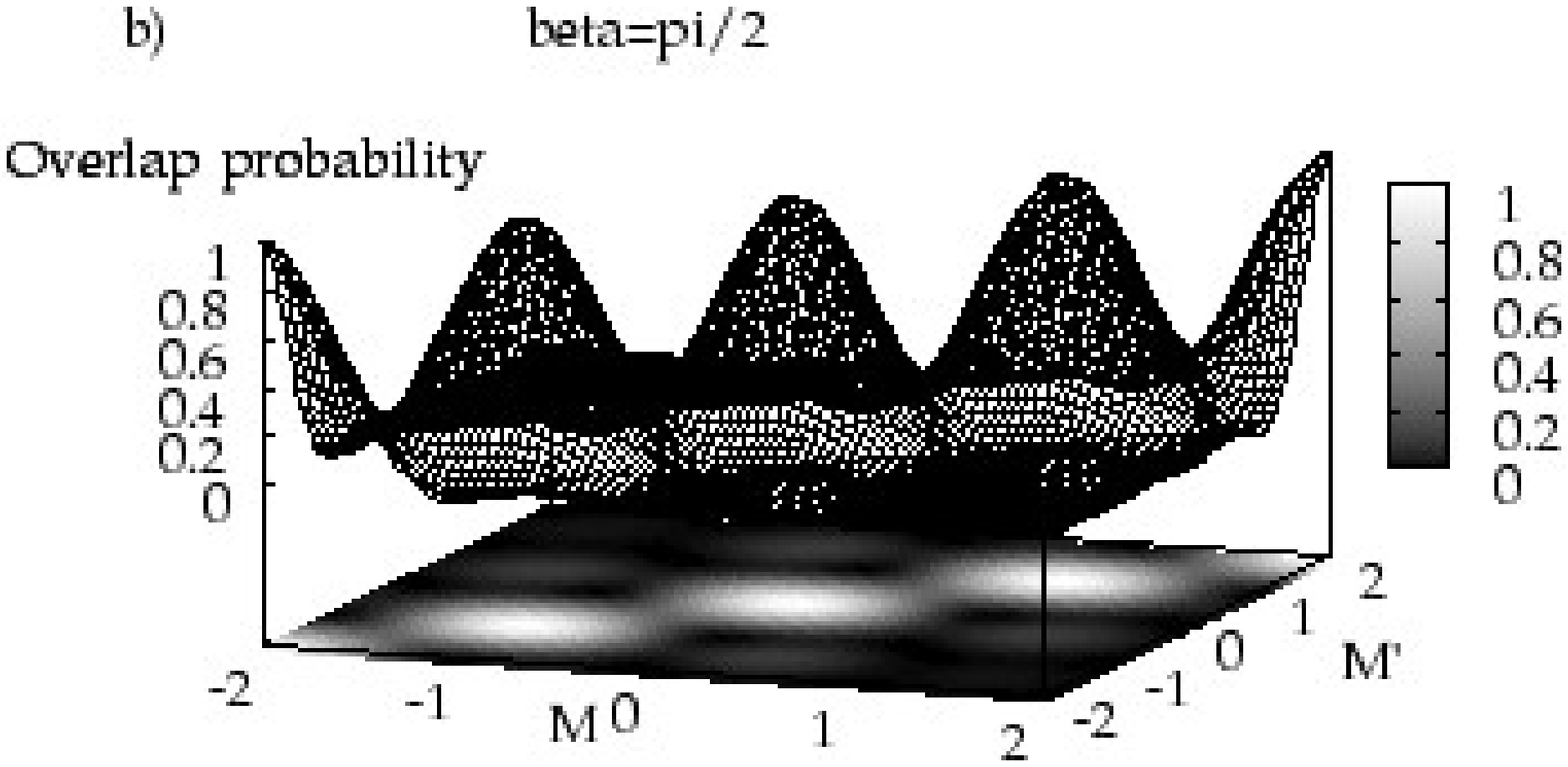}
 \includegraphics[width=0.7\textwidth]{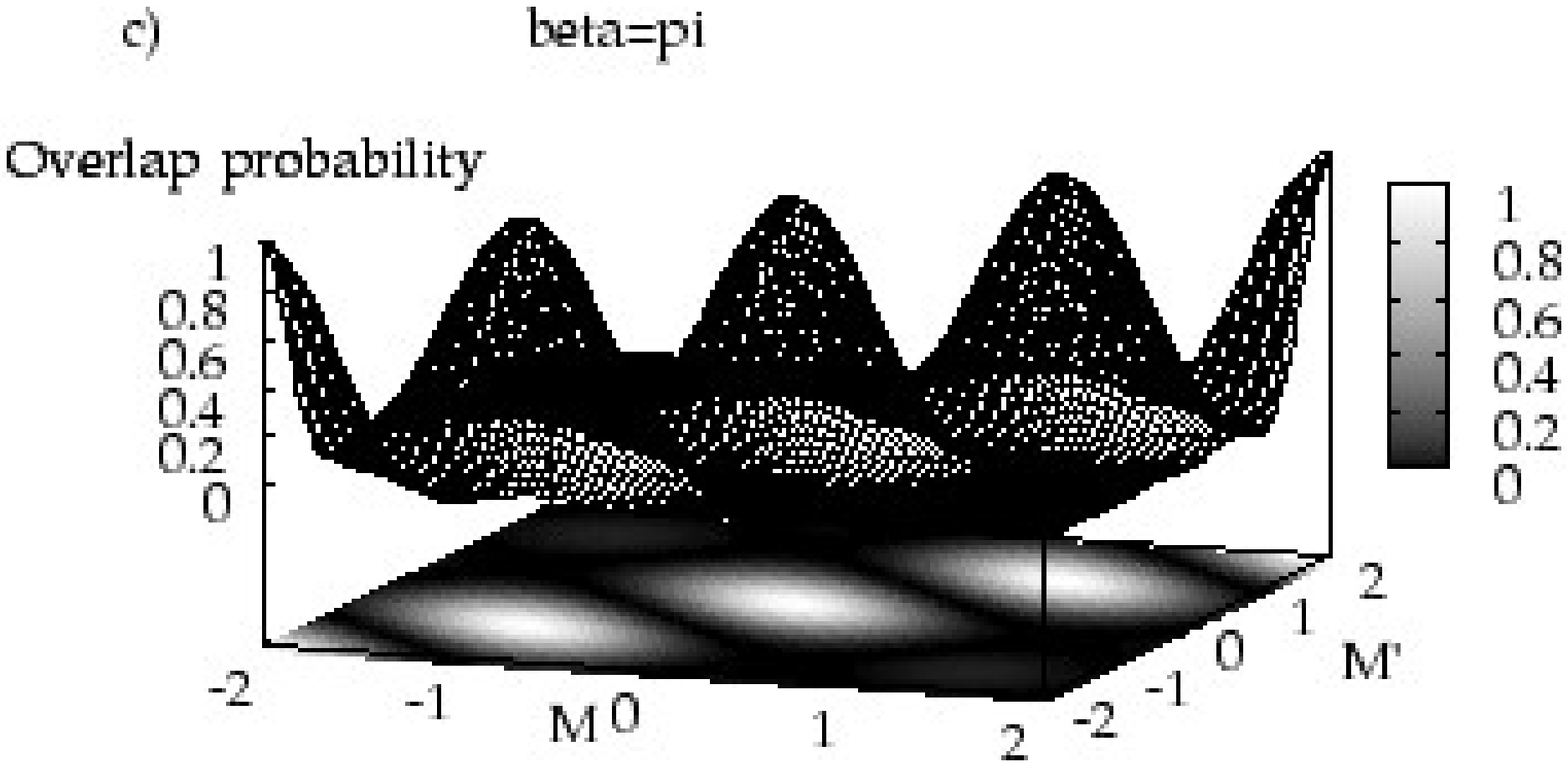}
 \end{center}
 \caption{\label{fig:genoverlap}Plot of the overlap probability for two general OAM states
 $|M \rangle$ and $|M' \rangle$ for three different values of the difference in the orientation $\beta$.
 a) $\beta=0$: Along parallel diagonal lines, for $M-M' = K \in \mathbb{Z}$ the overlap probability is independent of the particular values $M$ and $M'$.  b) $\beta=\pi/2$: The overlap probability generally depends on the fractional values, but does not reach zero for $M = M'$. c) $\beta = \pi$: For $M = M'$ and $\mu = \mu' = 1/2$, the
 overlap probability reaches zero.}
 \end{figure}

The study of high-dimensional two-photon entanglement reported in \cite{oemrawsingh+:prl92:2004,aiello+:pra72:2005,oemrawsingh+:prl95:2005} makes use of the fact that states with the same $M$ but with a $\pi$ radian difference in the orientation angle are orthogonal. With the expression for fractional states in Eq. (\ref{eq:fracstatedef})  we are able to calculate the overlap for two general states 
$| M(\alpha) \rangle$ and $| M'(\alpha') \rangle$. With help of the unitary operator $\hat{U}_m(-\alpha')$  we can set one orientation to zero and the other to the difference in the orientation $\beta = \alpha \oplus (- \alpha')$:
\begin{equation}
\label{eq:rotoverlap}
\begin{split}
\langle M'(\alpha') | M(\alpha) \rangle & = \langle M'(\alpha') | \hat{U}^\dagger_{m'}(-\alpha') 
\hat{U}_{m'}(-\alpha') | M(\alpha) \rangle =  \exp[\rmi(m-m')\alpha'] \langle M'(0) | M(\beta) \rangle, \\
&  = \exp[\rmi(m-m')\alpha'] \frac{\exp(-\rmi \mu \beta)}{2L+1} \sum_{n=0}^{2L} \exp[\rmi (M-M')\theta_n] \exp[\rmi \mu 2\pi f_\beta(\theta_n)],
\end{split}
\end{equation}
where we have used the fact that \{$| \theta_n \rangle\}_{n= 0,\dots,2L}$ is an orthonormal set. 
The moduli of the rotated and unrotated overlaps are the same so we proceed in calculating
$ \langle M'(0) | M(\beta) \rangle$ only. The sum over $n$ in  Eq. (\ref{eq:rotoverlap}) can be split in two parts by introducing an index $N$ with $\theta_N < \beta \leq \theta_{N+1}$ corresponding to the different cases in the definition of $f_\beta$ in Eq. (\ref{eq:ffunction}):
\begin{equation}
\langle M'(0) | M(\beta) \rangle = \frac{\exp(-\rmi\mu\beta)}{2L+1} \left[ \exp(\rmi \mu 2\pi) 
\sum_{n=0}^N \exp[\rmi (M-M') \theta_n] +
\sum_{n=N+1}^{2L} \exp[\rmi (M-M')\theta_n \right].
\end{equation}
We find that, on substituting $\theta_n = \theta_0+2\pi n /(L+1)$ into the expression for $\langle M'(0) | M(\beta) \rangle$, it is possible to evaluate the sums using geometric progression. The overlap is a physical result and we can take the limit of $L \to \infty$ by  expanding
the exponentials in the denominators. In this limit the set of angles $\{\theta_n\}_{n = 0,\dots,2L}$ 
becomes dense and we can write $\lim_{L \to \infty} \frac{2\pi(N+1)}{2L+1} = \lim_{L \to \infty} \theta_{N+1}  = \beta$:
\begin{equation}
\label{eq:genoverlap}
\begin{split}
\langle M'(0) | M(\beta) \rangle & = \exp(-\rmi\mu\beta) \frac{\rmi \exp[\rmi(M-M')\theta_0]}{2\pi(M-M')} \left\{ \exp(\rmi \mu 2\pi)
\left( 1 - \exp[\rmi(M-M')\beta] \right) \right. \\ 
& \left. + \left( \exp[\rmi(M-M')\beta] - \exp[\rmi (M-M') 2\pi] \right) \right\}.
\end{split}
\end{equation}
From this expression it is possible to calculate the modulus square of the overlap $\langle M'(0) | M(\beta) \rangle$ with help of several trigonometric identities: 
\begin{equation}
\begin{split}
| \langle M'(0) | M(\beta) \rangle |^2  & = \frac{1}{(M-M')^2 \pi^2} \left[ \sin^2(M \pi) + \sin^2(M' \pi)  \right. \\
& - \left. 2 \cos[(M-M')(\pi-\beta)] \sin(M \pi) \sin^{\vphantom{2}}(M' \pi) \right].
\end{split}
\end{equation}
This overlap probability is plotted for different values of $\beta$ in figure \ref{fig:genoverlap}. 
For $\beta = 0$ the plot shows that there are parallel lines of equal overlap probability for the diagonals with 
$M-M'=K \in \mathbb{R}$. The overlap probability along these lines is given by $\sin^2(\pi K)/(\pi^2 K^2)$. For $\beta \neq 0$ the overlap probability along the diagonal lines $M-M'=K$ becomes dependent on
$M$ and $M'$. In particular for $\beta = \pi$ and $M=M'$ the overlap probability reaches zero for 
$\mu=\mu'=1/2$. This was first reported in \cite{oemrawsingh+:prl92:2004} and, if we specialise the expression for the overlap probability in Eq. (\ref{eq:genoverlap}) to $M=M'$, we obtain an identical expression for the overlap probability $\langle M(0) | M'(\beta) \rangle$ which only depends on the
fractional part $\mu$ and the orientation $\beta$:
\begin{equation}
| \langle M(0) | M(\beta) \rangle |^2 = \left( 1 - \frac{\beta}{\pi} \right)^2 \sin^2(\mu \pi) + 
\cos^2(\mu \pi).
\end{equation}
This property has been used in an experimental detection of entanglement for fractional orbital angular momentum \cite{oemrawsingh+:prl95:2005} and in a proposed experiment to test for non-locality of
orbital angular momentum states \cite{aiello+:pra72:2005}.

 \subsection{Angular momentum distribution for fractional states}
 
\begin{figure}
\begin{center}
\includegraphics[width=0.7\textwidth]{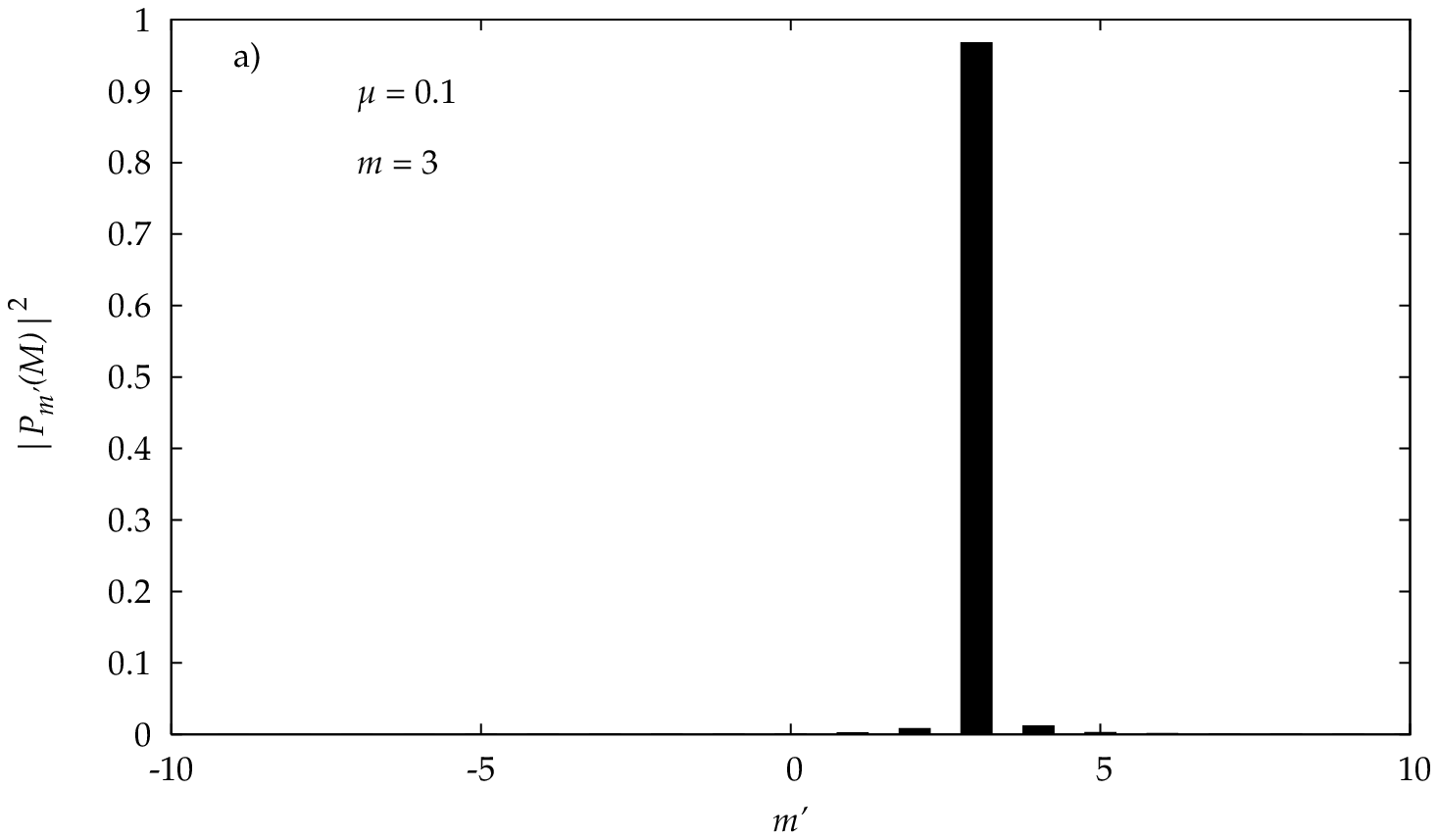}
\includegraphics[width=0.7\textwidth]{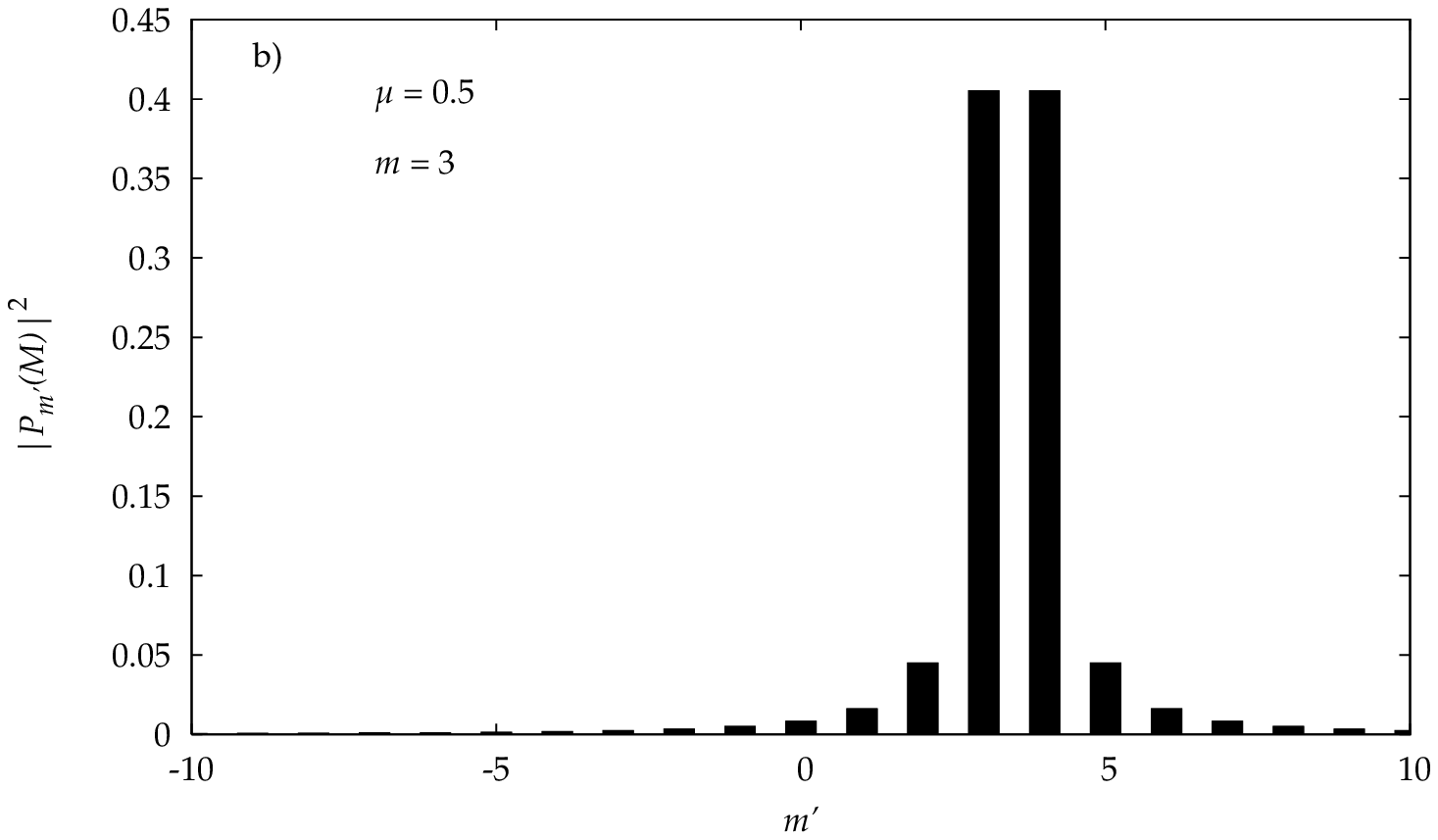}
\includegraphics[width=0.7\textwidth]{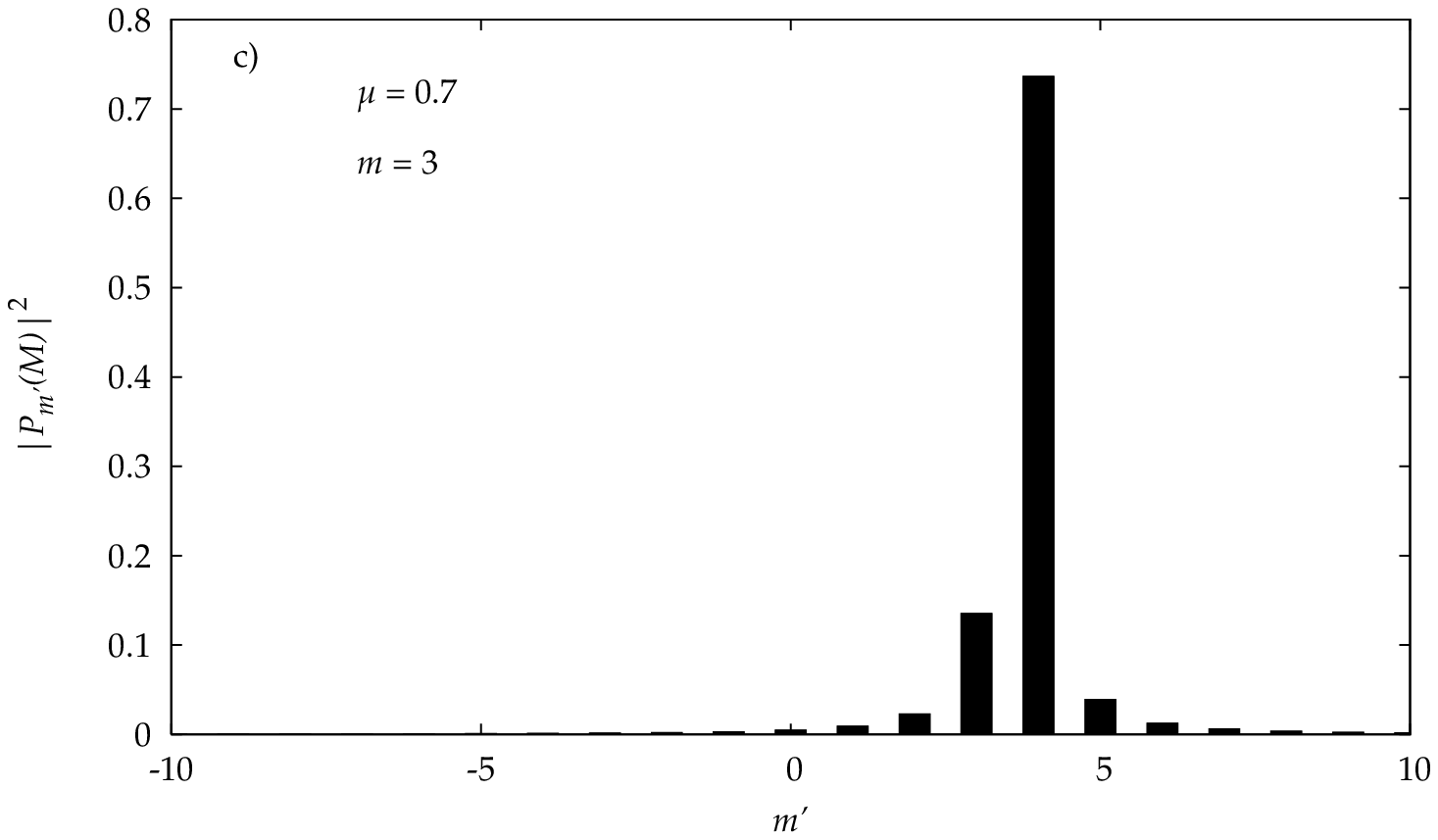}
\caption{\label{fig:oamdist} Plot of the OAM distribution of fractional states for different fractional values $\mu$. The distribution
has a peak at the nearest integer to $M$. Choosing a different integer value $m'$ shifts the distribution uniformly by $m' -m$, such that the distribution is peaked around $m'$ instead of $m$.  In b), for $\mu = 0.5$ this results in two peaks of equal height at the two neighbouring integers. The spread in the distribution is determined by the fractional value $\mu$. }
\end{center}
\end{figure}

We can decompose a fractional OAM state into the basis of integer OAM states. The amplitudes of the decomposition are given by the expression for the general overlap (\ref{eq:genoverlap}), where we choose $M'=m' \in \mathbb{Z}$, and we obtain:
\begin{equation}
\label{eq:fracoamdist} 
c_{m'}[M(\beta)]  = \langle m' | M(\beta) \rangle = \exp(-\rmi \mu \beta)\frac{\rmi \exp[\rmi(M-m')\theta_0]}{2\pi(M-m')} \left[ \exp[\rmi(m-m')\beta]
\left( 1 - \exp(\rmi \mu 2\pi) \right) \right].
\end{equation}
Only the complex argument of the probability amplitudes $c_{m'}[M(\beta)]$ depends on the relative orientation;
the probabilities $P_{m'}(M)$, given by the modulus square of the probability amplitudes, are independent of $\beta$:
\begin{equation}
P_{m'}(M) = |c_{m'}[M(\beta)]|^2 = \frac{1- \cos(2\mu\pi)}{(M-m')^2 2\pi^2} = \frac{\sin^2(\mu \pi)}{(M-m')^2\pi^2}.
\end{equation}
Integer OAM states form an orthonormal basis, that is they obey $\langle m | m' \rangle = 
\delta_{mm'}$ for $m,m' \in \mathbb{Z}$. This particular result is recovered for the OAM probabilities 
$P_{m'}(M)$ from Eq. (\ref{eq:fracoamdist}). For $M = m \in \mathbb{Z}$ and $m \neq m'$ the fractional part $\mu$ is zero resulting in a vanishing OAM probability. For $M = m'$, we can determine the value of $P_{m'}(m')$ by the limiting procedure:
\begin{equation}
P_{m'}(m) = \lim_{\mu \to 0} P_{m'}(m + \mu) = \frac{1}{2\pi^2} \lim_{\mu \to 0} \frac{1 - \cos(2\mu\pi)}
{(m + \mu - m')^2} = \delta_{mm'}.
\end{equation}
For integer values of $M$ the OAM distribution is thus singular, consisting of a single 
non-vanishing probability at $M = m'$. For fractional values of $M$, however, the probabilities are peaked around the nearest integer to $M$, as can be seen in figure \ref{fig:oamdist}.

\begin{figure}
\begin{center}
\includegraphics[width=0.7\textwidth]{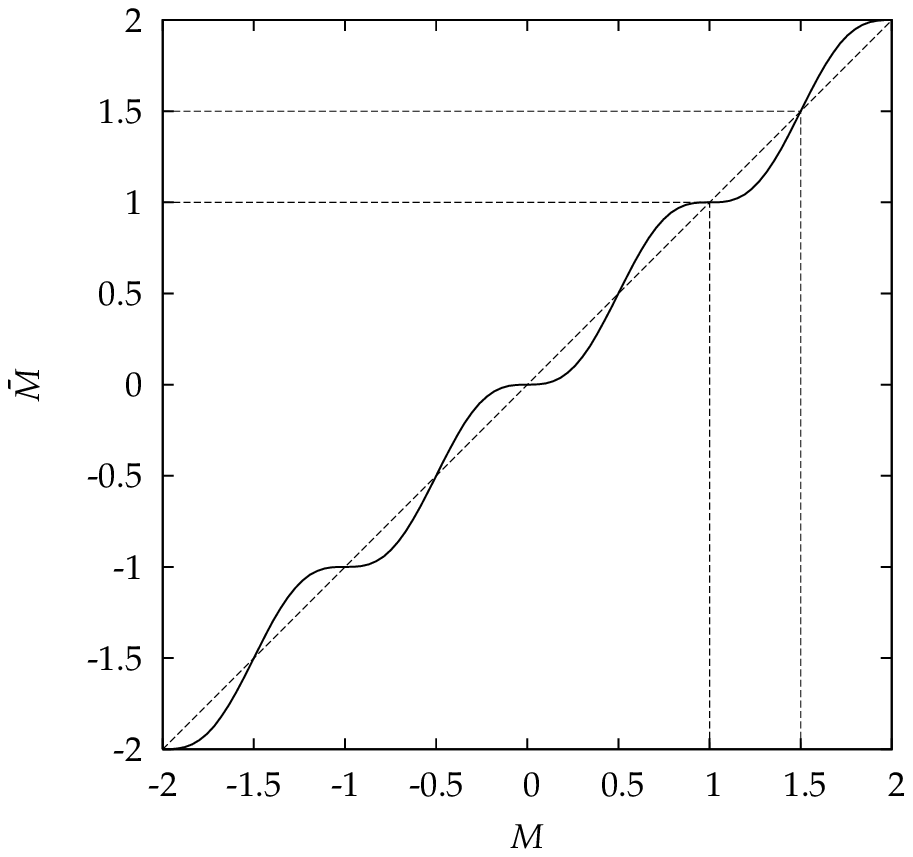}
\caption{\label{fig:oammean} Plot of the OAM mean value for fractional OAM states. Only at integer and half integer values of $M$ is the OAM mean
$\bar{M}$ equal to $M$. In general the relation between $M$ and $\bar{M}$ is given in Eq. 
(\ref{eq:oammean}).}
\end{center}
\end{figure}

Owing to the completeness of the OAM basis states $| m' \rangle$,
the probabilities $P_{m'}(M)$ sum to unity.
The relevant summation can be executed using the
contour integration method \cite{stephenson+:cup:1993}. In a similar way the mean value of the OAM can be calculated:
\begin{equation}
\label{eq:oammean}
\bar{M} = \sum_{m'=-\infty}^\infty m' P_{m'}(M) = \frac{1-\cos(2 M \pi)}{2\pi^2} \sum_{m'=-\infty}^\infty 
\frac{m'}{(M-m')^2} = M - \frac{\sin(2 M \pi)}{2\pi},
\end{equation}
where we have used $\cos(2 \mu \pi) = \cos(2 M \pi)$.
The result is an OAM mean that is equal to $M$ only at integer and half-integer values of $M$ (see figure \ref{fig:oammean}). This expression was first reported by Berry in \cite{leach+:njp6:2004} where it was also confirmed by experiments.
The OAM variance, however, tends to infinity.
From a more general point of view it is not surprising that the OAM variance is divergent.
The discontinuity in the phase leads also to a discontinuity in the angle wavefunction. For such a discontinuous wavefunction the mean square OAM will be divergent
\cite{pegg+:njp7:2005}. 

\section{Propagation of a fractional beam}

The decomposition of a fractional state in OAM eigenstates can be applied directly to
the decomposition of a light mode emerging from a fractional phase step into OAM eigenmodes. By using a decomposition into Bessel beams rather than Laguerre-Gaussian beams we can switch conveniently from a non-paraxial solution to a paraxial solution by changing the 
propagation factor in the Bessel beam. Berry has reported the propagation of a plane wave with a fractional OAM imprinted on it \cite{berry:joa6:2004}. Here, we find many of the same features for a beam of finite transverse extent.

For the decomposition we consider a Gaussian input beam with a finite beam width $w_0$.  The beam emerging from spiral phase plate can thus be written
as 
\begin{equation}
\label{eq:gausszero}
\Psi(\rho, \varphi, z=0) = \exp \left( - \frac{\rho^2}{2 w_0^2} \right) \exp(\rmi M \varphi).
\end{equation}
A further advantage of Bessel functions as solutions to the exact and paraxial wave equation is that the
dependence on the radial and longitudinal coordinates factorises. We can therefore express the
Gaussian beam directly behind the phase step as a superposition of Bessel beams with different transverse wavenumbers $\kappa$.  The propagated wave is obtained by multiplying each term in the superposition by the appropriate propagation factor for the exact or paraxial theory. The superposition of Bessel beams is specific for every integer value of the orbital angular momentum and the decomposition of the wave behind a fractional phase step requires a summation over integer values $m$ weighted by the coefficients $c_{m'}[M(\beta)]$:
\begin{equation}
\label{eq:wavezero}
\Psi(\rho, \varphi, z=0) = \sum_{m' = -\infty}^\infty c_{m'}[M(\beta)] \exp(\rmi m' \varphi) \int \rmd \kappa \: 
d_{m'}(\kappa) J_{m'}(\kappa \rho).
\end{equation} 
The coefficients $c_{m'}[M(\beta)]$ are given by the overlap between the integer and fractional OAM states $\langle m' | M(\beta) \rangle$ (see Eq. (\ref{eq:fracoamdist})). 
This follows on multiplying both expressions for $\Psi(\rho, \varphi, z=0)$ in Eqs.
(\ref{eq:gausszero}) and (\ref{eq:wavezero}) by $\exp(-\rmi m'' \phi)$ and integration over $\varphi$. 
The radial part contains the function $d_{m'}(\kappa)$ which is determined by the integral
\begin{equation}
\label{eq:besseldecomp}
 \exp \left( - \frac{\rho^2}{2 w_0^2} \right) = \int \rmd \kappa \: 
d_{m'}(\kappa) J_{m'}(\kappa \rho).
\end{equation}
On using the Fourier-Bessel theorem \cite{gray+:mac:1895}, the function $d_{m'}(\kappa)$ can be found to be:
\begin{equation}
\label{eq:coefintegral}
d_{m'}(\kappa) = \kappa \int_0^\infty \rmd \rho \: \rho J_{m'} (\kappa \rho) \exp
\left( -\frac{\rho^2}{2 w_0^2} \right).
\end{equation}
The solution of this integral can be written in terms of modified Bessel functions $I_\nu$ with
a half-integer order $\nu$ \cite{gradshteyn+:ap:2000}. For all positive half-integer orders the modified Bessel functions increase exponentially. But this behaviour changes for $\nu = -\frac{1}{2}$ and $\nu = \frac{3}{2}$ which renders the integral expression for the coefficients in terms of
modified Bessel functions invalid for $m' \leq - 2$. We therefore have to evaluate the integral in 
 Eq. (\ref{eq:coefintegral}) for negative values of $m'$ separately by using $J_{-m'}(\kappa \rho)
 = (-1)^{m'} J_{m'}(\kappa \rho)$ \cite{stephenson+:cup:1993}. The coefficients in Eq. 
 (\ref{eq:coefintegral}) may thus be written as
 \begin{equation}
 \label{eq:coefffinal}
 d_{m'}(\kappa)  = \kappa^2 \frac{\sqrt{\pi}}{8} \left( 2w_0^2 \right)^{\frac{3}{2}} \exp \left( -\frac{\kappa^2 w_0^2}{4} \right) 
\left[ I_{\frac{|m'|-1}{2}} \left(-\frac{\kappa^2 w_0^2}{4} \right) - I_{\frac{|m'|+1}{2}} \left(-\frac{\kappa^2 w_0^2}{4} \right) \right] \times \left\{ \begin{array}{ll} (-1)^{|m'|} & m' < 0, \\ 1 & m' \geq 0. \end{array} \right.
 \end{equation}

On substituting the coefficients in Eq. (\ref{eq:wavezero}) the wave emerging from a fractional phase step is determined at $z=0$. To calculate the propagated wave we have to add the
appropriate propagation factor in the integral in Eq. (\ref{eq:wavezero}). For the exact form
the propagation factor is given by $\exp(\rmi \sqrt{k^2 - \kappa^2}z)$ and the propagated wave reads as
\begin{equation}
\label{eq:exact}
\Psi(\rho, \varphi, z)  = \sum_{m' = -\infty}^\infty c_{m'}[M(\beta)] \exp(\rmi m' \varphi) 
\int_0^\infty \rmd \kappa \: d_{|m'|}(\kappa) J_{|m'|}(\kappa \rho) \exp(\rmi \sqrt{k^2 - \kappa^2}z).
\end{equation}
A distinction between negative and positive values of $m'$ is not necessary in this formulation as for
negative $m'$ the alternating factor $(-1)^{m'}$ is compensated by the alternating sign of the Bessel functions $J_{-m'}(\kappa \rho) = (-1)^{m'} J_{m'}(\kappa \rho)$. 
The integration boundaries are set from zero to infinity and this includes evanescent waves for 
$\kappa > k$. As pointed out in \cite{roux:oc223:2003}, the evanescent components are necessary to 
describe the singularity correctly. It is worth noting, however, that
the character of the integral changes at $\kappa = k$, where the square root turns imaginary and the
exponential changes from an oscillating to a decreasing behaviour. In the paraxial solution, where
the propagation factor is $\exp[ \rmi k (1 - 1/(2 \kappa^2) z] $, this distinction does not exist. The
$z$ dependent term $\exp(\rmi k z)$ in the propagation factor can be written in front of the 
integral, which gives for the paraxial solution
\begin{equation}
\label{eq:paraxial}
\psi(\rho, \varphi, z)  = \sum_{m' = -\infty}^\infty c_{m'}[M(\beta)] \exp(\rmi m' \varphi) \exp(\rmi k z) 
\int_0^\infty \rmd \kappa \: d_{|m'|}(\kappa) J_{|m'|}(\kappa \rho) \exp \left(- \rmi \frac{k} {\kappa^2} z
\right).
\end{equation}
The integrals for the exact and paraxial solution are solved numerically to give the intensity and the
phase profile of the propagated wave. 

\section{Numerical results}

\begin{figure}[p]
\begin{center}
\includegraphics[width=0.4\textwidth]{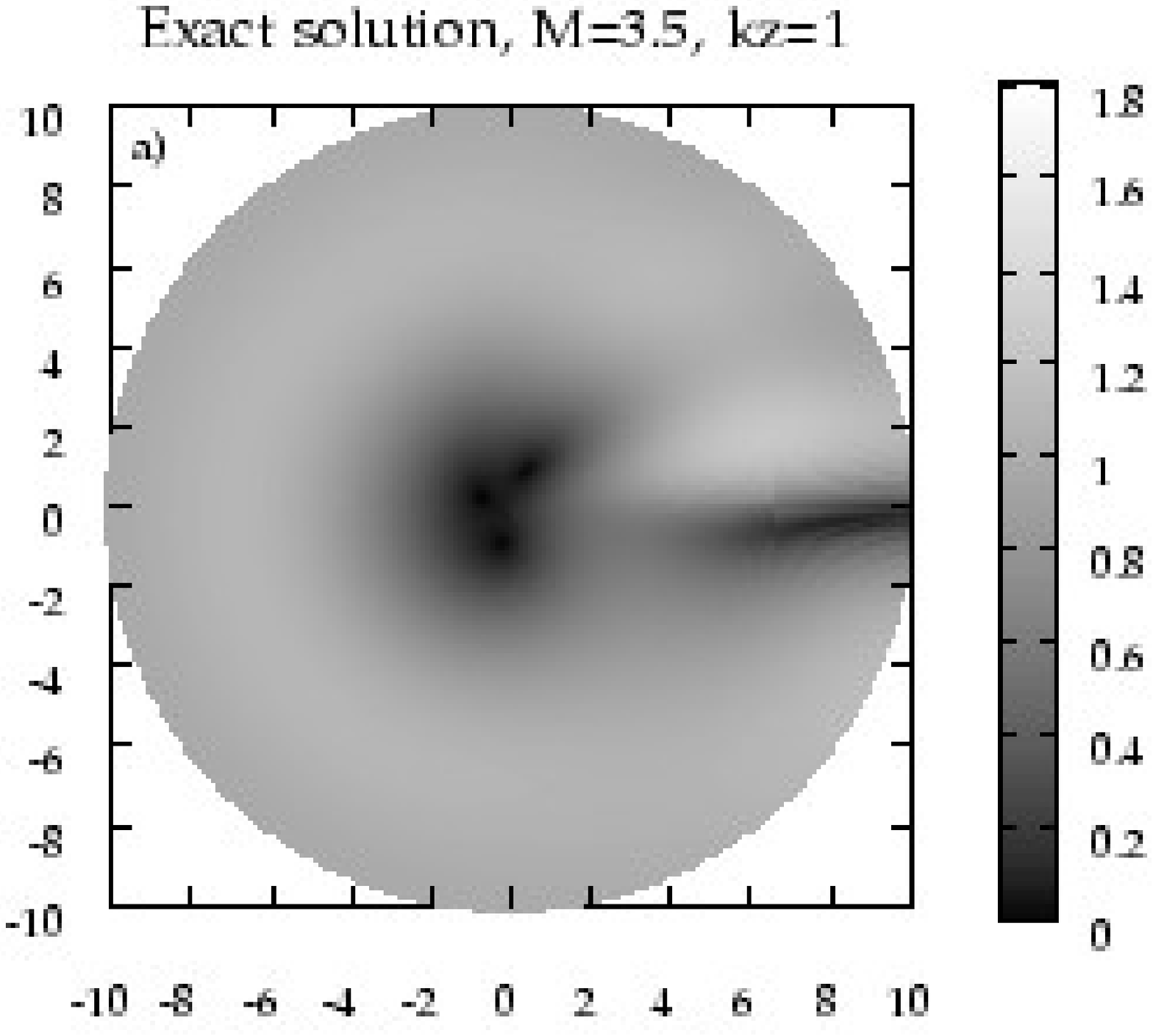}
\includegraphics[width=0.4\textwidth]{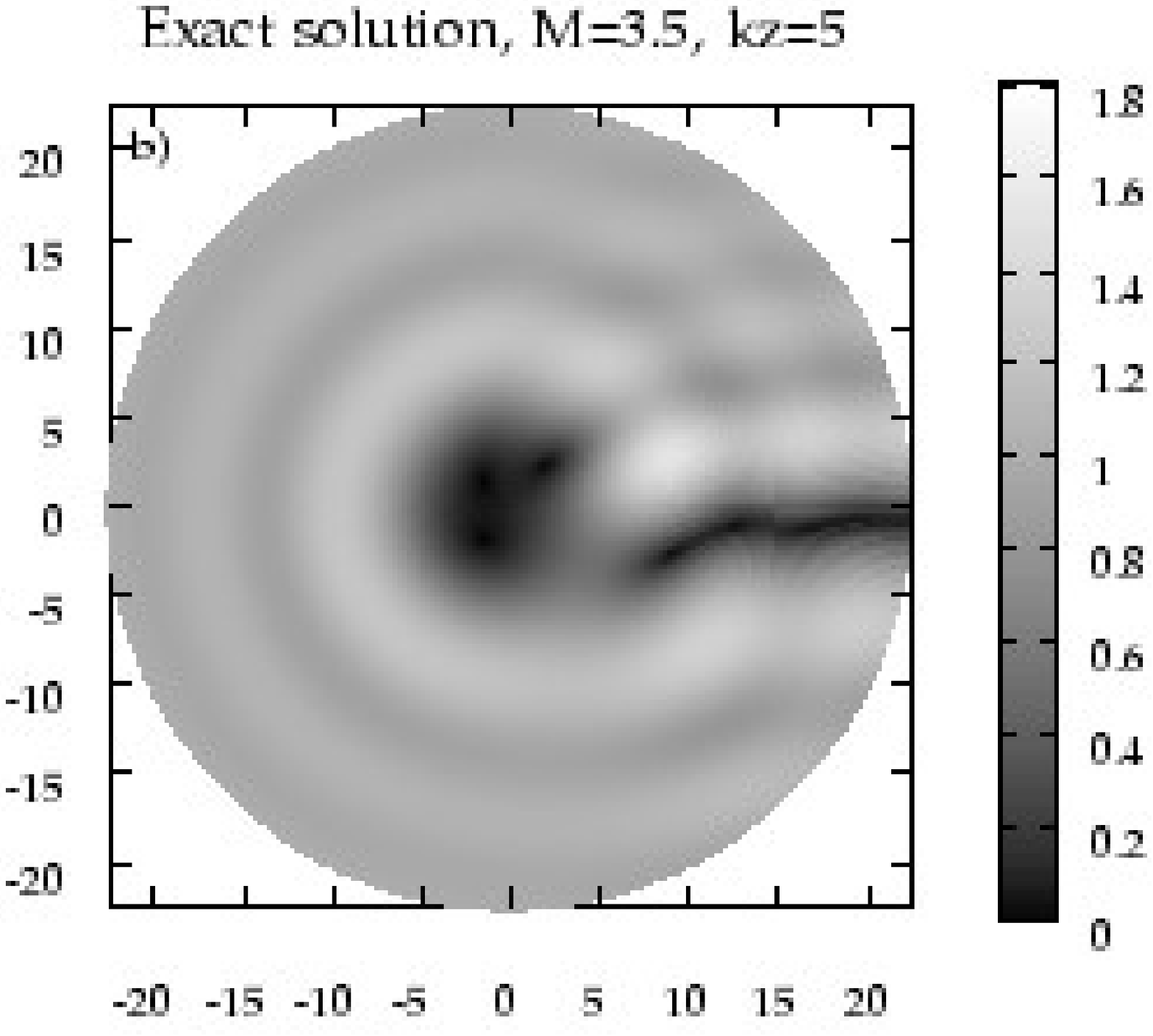} \\
\includegraphics[width=0.4\textwidth]{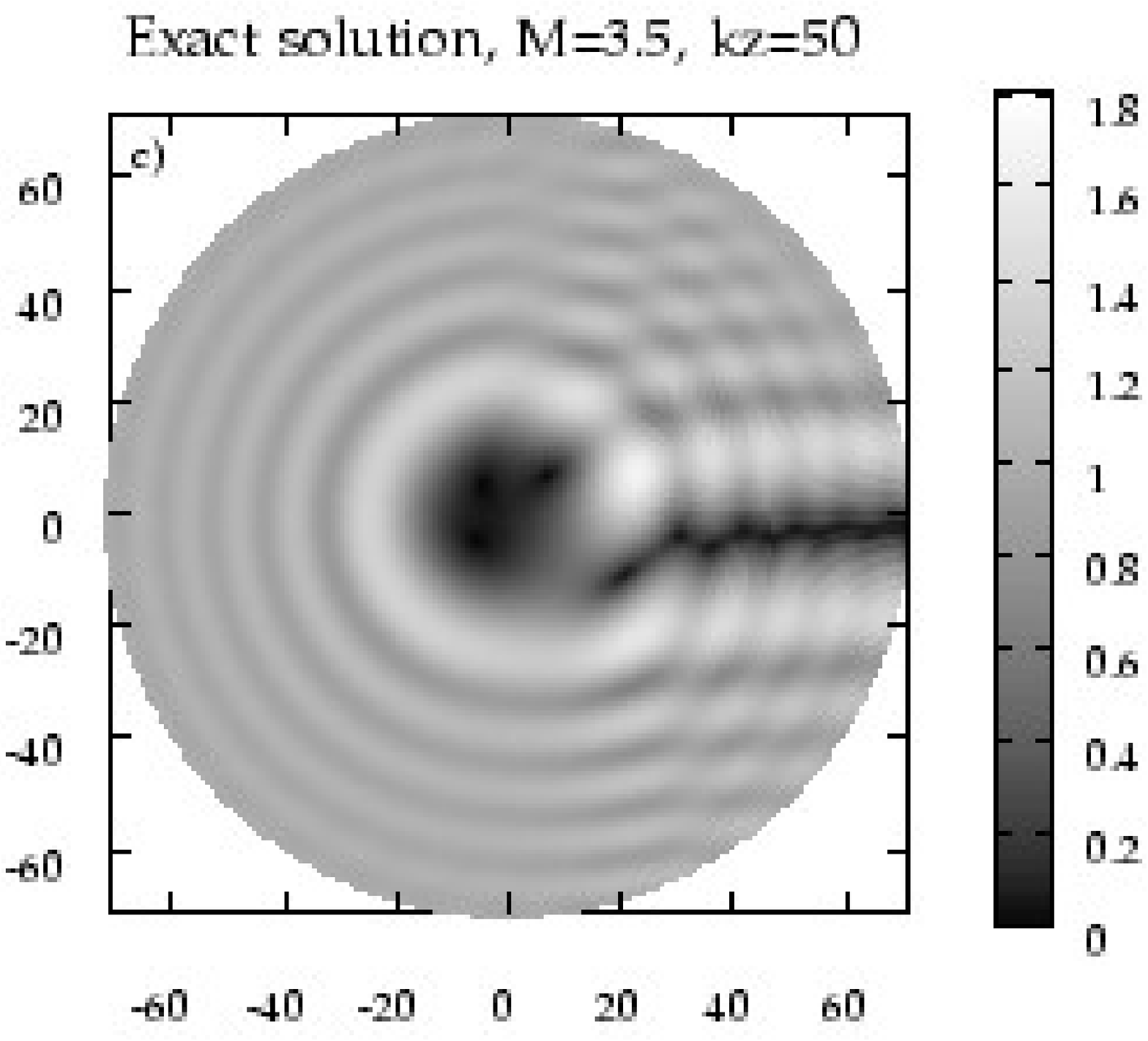} 
\includegraphics[width=0.4\textwidth]{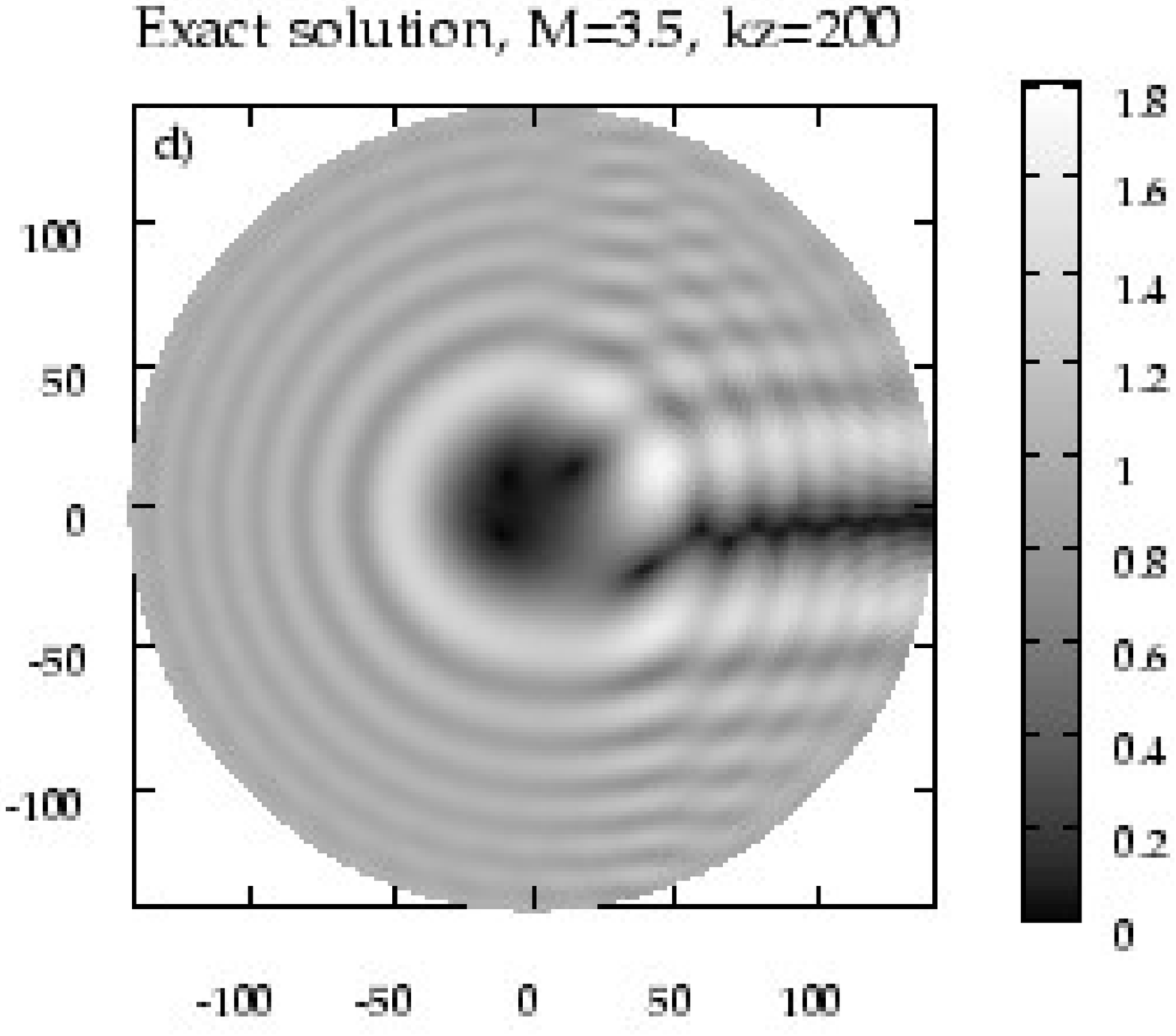} \\
\includegraphics[width=0.4\textwidth]{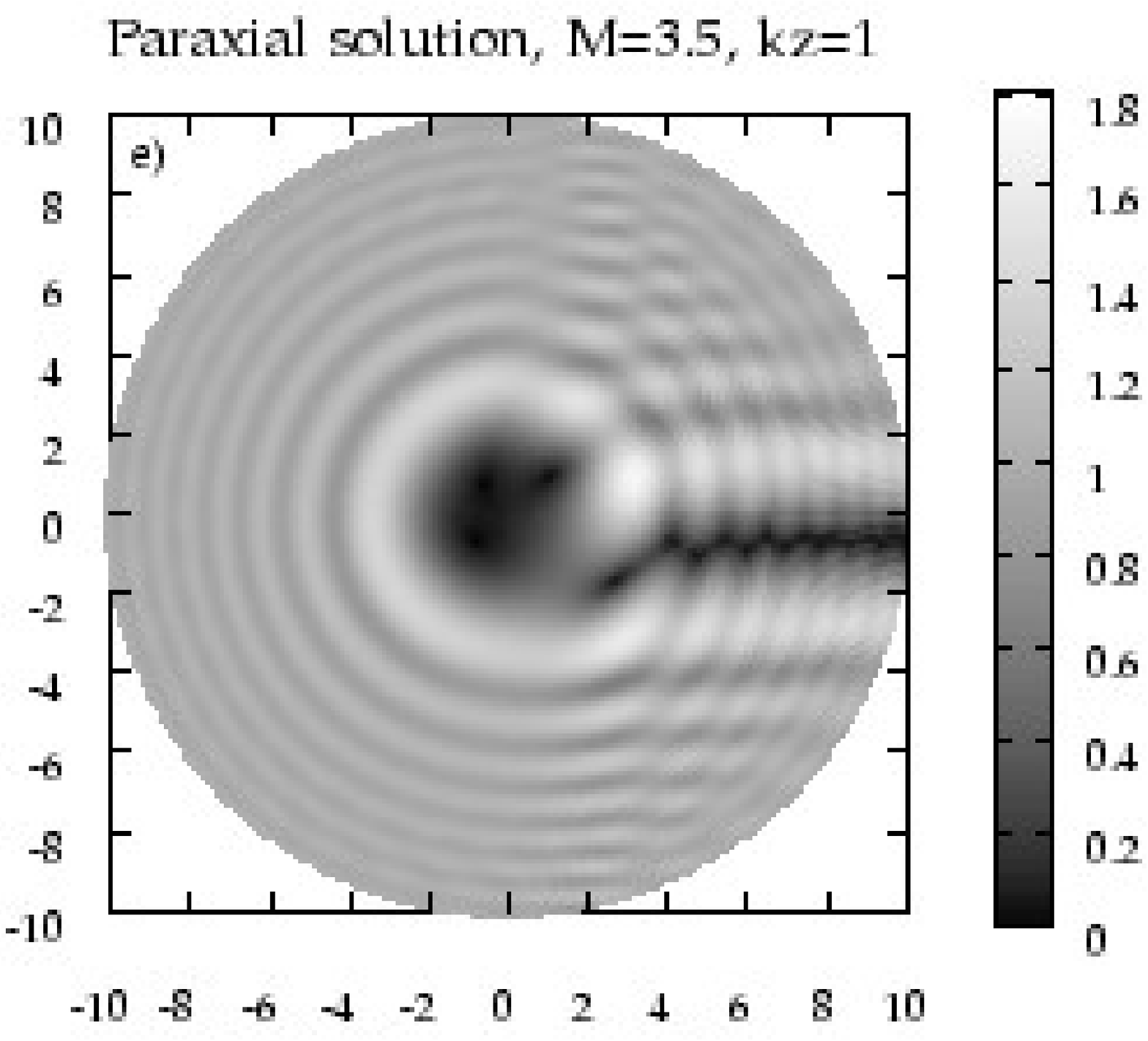}
\includegraphics[width=0.4\textwidth]{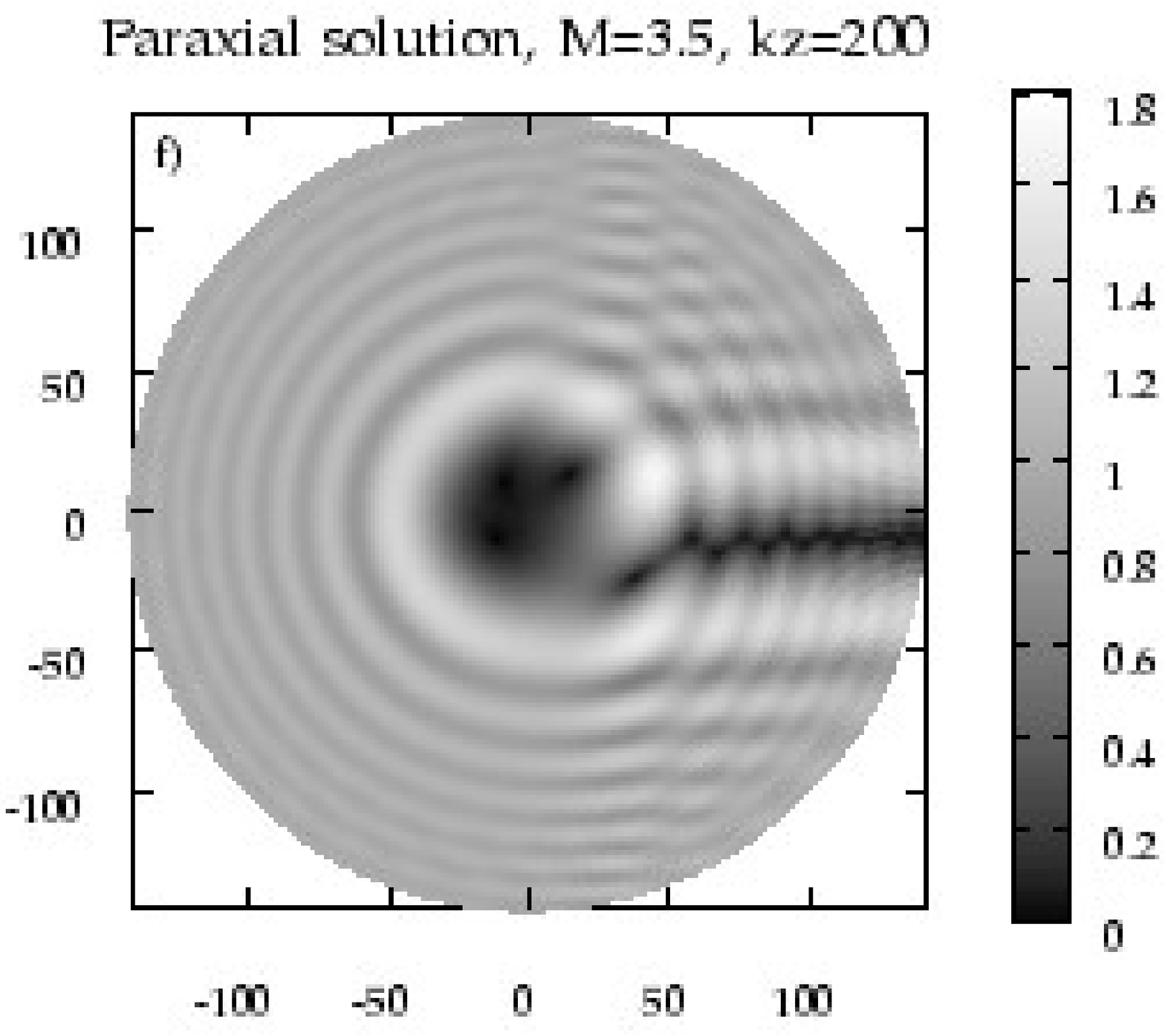}
\caption{\label{fig:insty35}Intensity profiles for $M=3.5$. The graphs a-d) show the
exact solution for the propagation distance $kz=1,5,50,200$ respectively. The graphs e) and f) show
the paraxial solution for the distances $kz=1$ and $kz=200$. For fractional phase steps with 
$\mu = 1/2$ the intensity along the radial line drops to zero.}
\end{center}
\end{figure}

\begin{figure}[p]
\begin{center}
\includegraphics[width=0.4\textwidth]{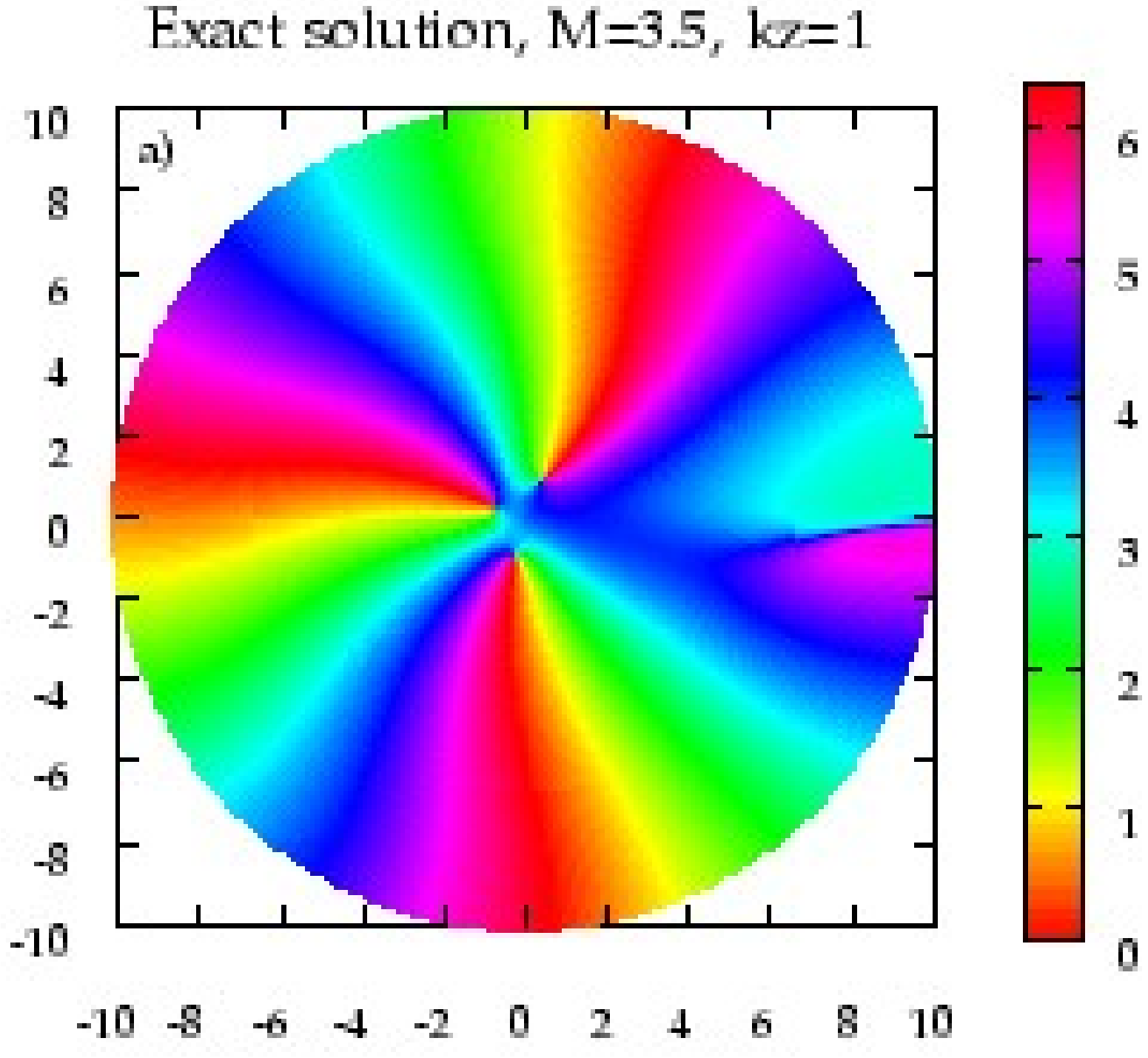}
\includegraphics[width=0.4\textwidth]{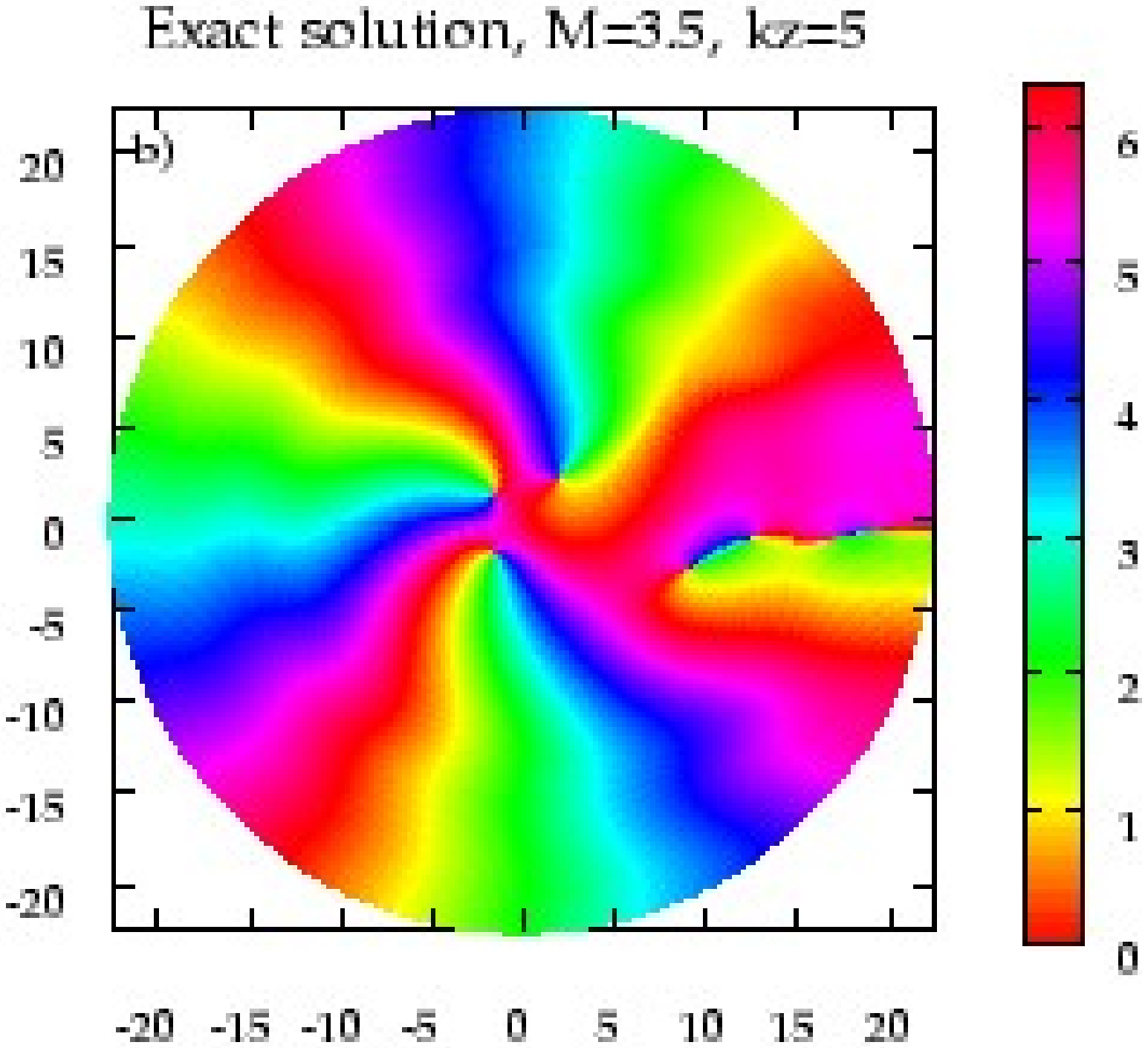} \\
\includegraphics[width=0.4\textwidth]{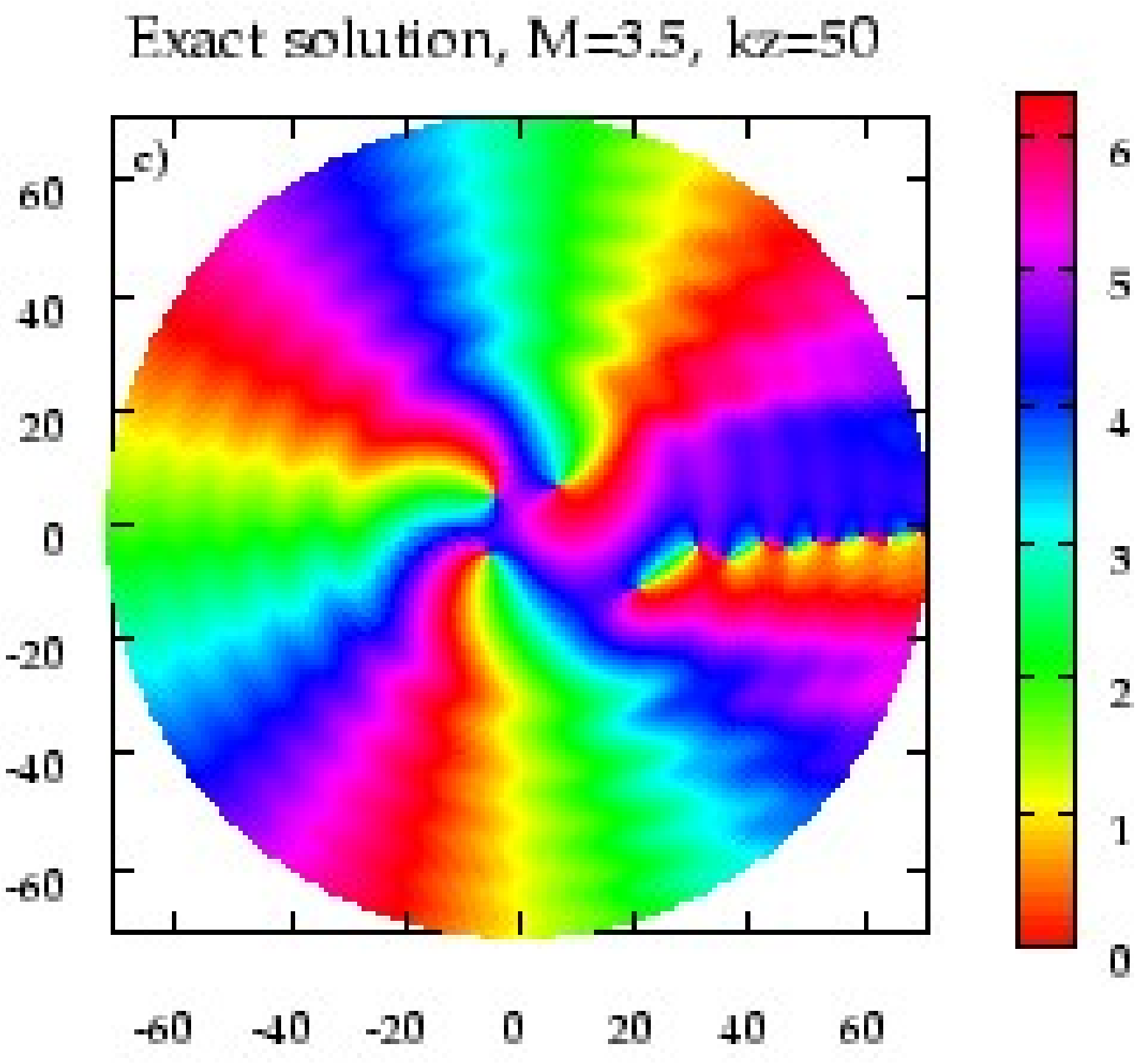} 
\includegraphics[width=0.4\textwidth]{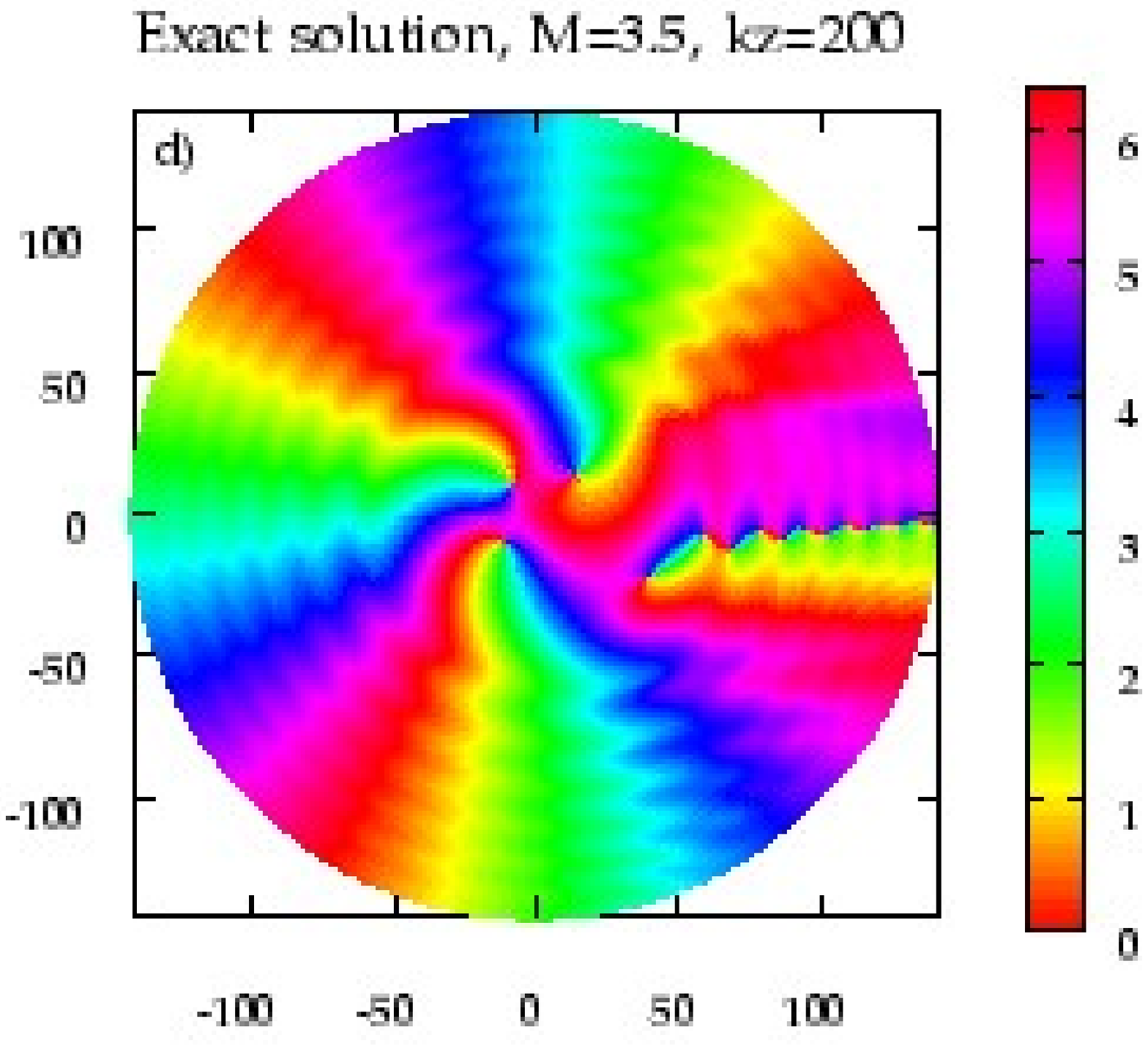} \\
\includegraphics[width=0.4\textwidth]{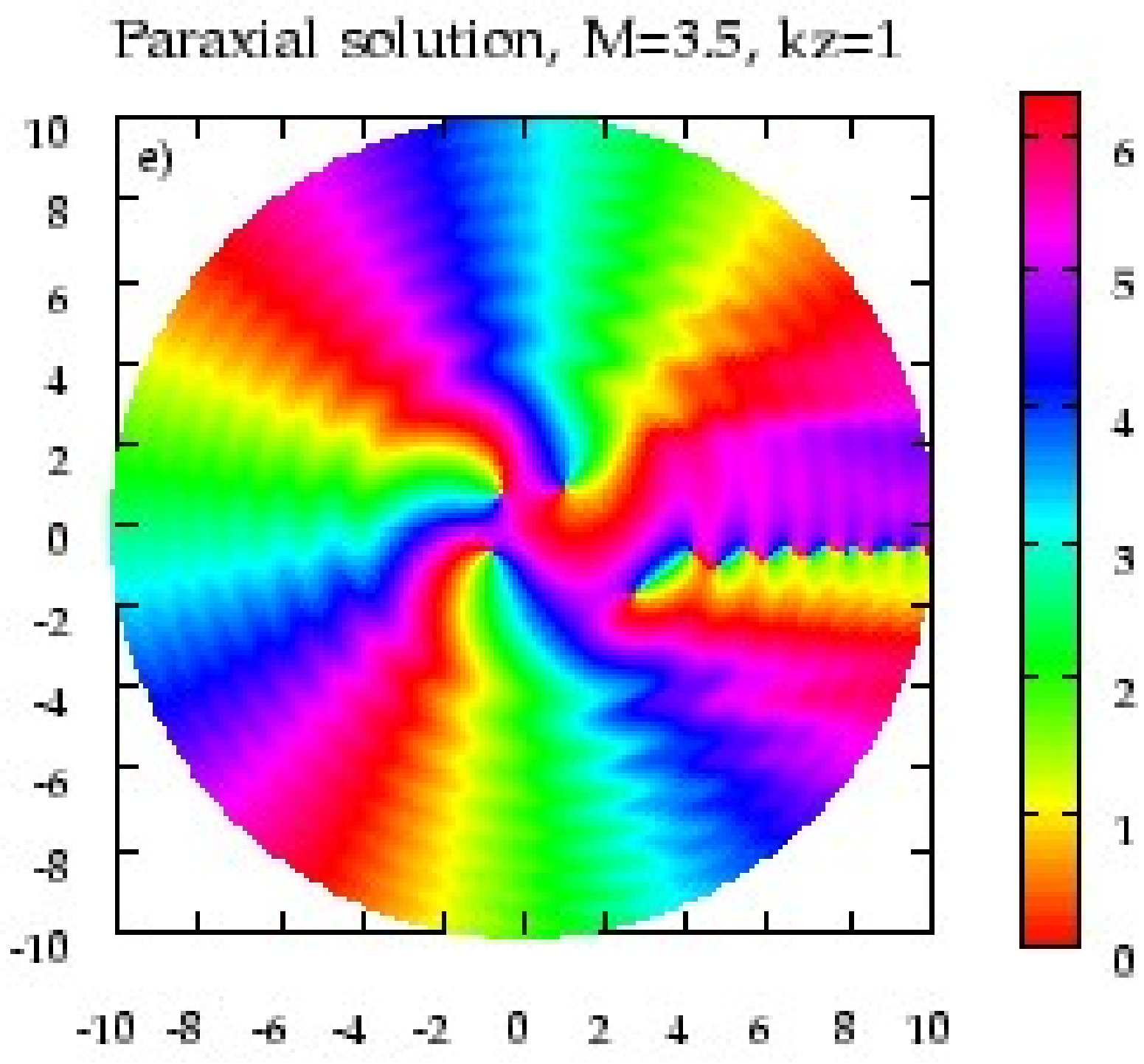}
\includegraphics[width=0.4\textwidth]{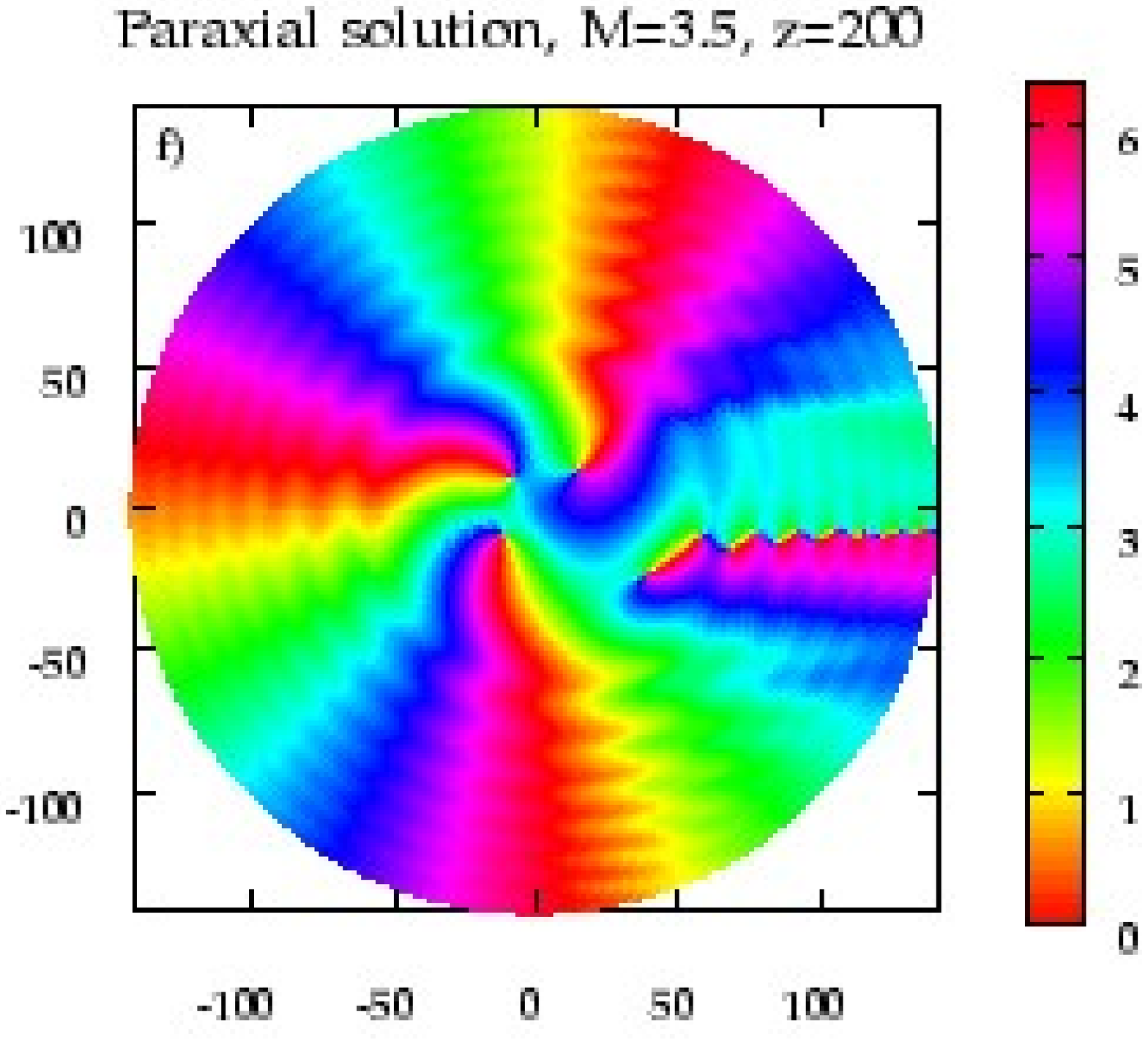}
\caption{\label{fig:phase35}Phase profiles for $M=3.5$. The graphs a-d) show the
exact solution for the propagation distance $kz=1,5,50,200$ respectively. The graphs e) and f) show
the paraxial solution for the distances $kz=1$ and $kz=200$. The phase profiles show the creation
of additional vortices on propagation.}
\end{center}
\end{figure}

For the numerical calculation we only take eigenmodes into account for which the corresponding 
OAM probability $|c_{m'}[M(\beta)]|^2$ is larger than $10^{-4}$. The widest spread in the OAM spectrum occurs for half odd-integer values of $M$ and for these the restriction to 
$|c_{m'}[M(\beta)]|^2 > 10^{-4}$ amounts to 66 integer eigenmodes contributing to the decomposition.
The phase and intensity profiles are calculated for the
fractional phase step $M = 3.5$, because for half odd-integer phase steps we expect to find the interesting phenomenon of the formation of a chain of vortices in the line of low intensity created by the phase discontinuity \cite{berry:joa6:2004,leach+:njp6:2004}. We give the results for both the exact solution and in the paraxial approximation. We choose
a beam width of $w_0 = 10000$ $\lambda / (2\pi)$, which corresponds to a beamwaist of about one millimeter for visible light. For the
paraxial solution and the chosen beam width  the intensity profile remains unchanged on propagation and the phase profile simply rotates around the propagation axis. This
is why we present four profiles for the exact solution at the propagation distances $kz = 1,5,50,200$ and
two for the paraxial solution at $kz=1,200$. The results are presented in two series of graphs in
figures \ref{fig:insty35} and \ref{fig:phase35}. In the intensity profiles the
most obvious feature is the radial line of zero intensity. For a value of $M$ with a fractional part different from $1/2$ the intensity will be low but not equal to zero along this line. The other
important feature in the intensity profiles is the number and position of the spots of
zero intensity in the centre of the beam. The  graph for $M =3.5$ shows
$3$ such spots. These spots correspond to optical vortices and the number
of vortices is given by the modulus of the nearest integer to $M$. While a beam with integer
OAM $m \in \mathbb{Z}$ propagates with an optical vortex of charge $m$ on the axis,  beams with fractional OAM only show vortices with charge $\pm 1$. None of these
vortices is on the axis, but the whole central region has low intensity. Another prominent feature are
the diffraction fringes surrounding the axis; these arise due to diffraction of the initial singularity at the centre. The radial fringes in the direction of $\alpha$ are due to diffraction on the discontinuity in the spiral phase plate or hologram. The contrast in the fringes is highest for a fractional part $\mu = 1/2$.
It is diffraction, of course, which is responsible for breaking the initial rotational symmetry.

The phase profiles show clearly that the fractional strength vortex splits up in vortices with
strength $\pm 1$. The phase is represented in false colour, such that the colours for the phase
of $0$ and $2\pi$ are identical. A vortex corresponds to a point where
all colours meet and for a vortex of strength $\pm 1$ a circle surrounding the vortex goes through
the colour circle once. The sign of $\mp1$ determines if the colours run through the circle clockwise or
anti-clockwise. The most prominent feature in the phase profiles is the creation of additional vortices in the region of zero intensity for half odd-integer phase steps.
In figure (\ref{fig:phase35}a), for $kz=1$, one can see that there are no additional vortices, while in figure 
(\ref{fig:phase35}b) one pair of vortices has formed and another pair is in the process of forming. At greater propagating distances the series of vortex pairs moves closer together. From the different rotation sense for the colours it can be seen that the vortices in this series have alternating charge. The vortices annihilate each other differently depending on whether the phase step has a fractional part $\mu$ of just under $1/2$ or just over $1/2$ \cite{berry:joa6:2004}. This leads to the formation
of a new vortex or the destruction of an existing vortex for fractional phase steps with $\mu > 1/2$.
The phase pattern rotates on propagation. For a light beam with integer OAM this rotation would be uniform, but due to the presence of the imprinted phase discontinuity, which does not change under propagation,  the rotation of the phase pattern changes. If one follows for example the red line of zero phase roughly along the negative $y$-axis in figure \ref{fig:phase35}a, this line will rotate counter-clockwise on propagation and at some distance it will be oriented along the negative $x$ axis. Within the same propagation distance the line of zero phase, that is roughly oriented along the positive $y$ axis, will rotate through the phase discontinuity until it is oriented along the negative $y$ axis. Over the same propagation distance these two lines of zero phase cover different rotation angles. This shows that the phase discontinuity at the angle $\alpha$ causes the non-uniform rotation of the phase pattern.

On propagation fractional modes develop a number of features which can only be seen in the
exact solution. Although the paraxial solution approaches the exact solution for $z \to \infty$ the
paraxial solution does not show the changing position of the integer vortices or the formation and movement in the chain of vortices.

\section{Conclusion}
 
In this article we have derived a rigorous quantum formulation of fractional orbital angular momentum (OAM). This formalism is a generalisation of the quantum theory of rotation angles by Barnett and Pegg
\cite{barnettpegg:pra41:1990} to fractional values of the OAM. Fractional values of OAM require the introduction of a branch cut
in the angle representation of the OAM eigenfunctions. The orientation of the branch cut 
is an additional parameter in the description of fractional OAM. We have calculated
the overlap of two general fractional OAM states and we have found that for half odd-integer values of the OAM and a relative orientation of $\pi$ radian the 
overlap is zero and the corresponding states are orthogonal. 
This confirms earlier work on fractional OAM with half-integer spiral phase plates \cite{oemrawsingh+:prl92:2004}. 

We have applied our theory to the propagation of light beams with fractional OAM. For light carrying integer OAM a stable optical vortex of corresponding
integer strength forms on the propagation axes. For fractional OAM no
fractional strength vortices are stable on propagation, and instead a number of strength $\pm 1$
vortices are formed in the central region of the beam. Additionally, a line of low intensity is visible
corresponding to the orientation of the branch cut or the phase step discontinuity in the generating optical device. For half-integer phase steps a chain of additional vortices with alternating charge is formed in this region of low intensity. We have confirmed theoretical and experimental results
regarding the evolution of optical vortices in light beams with fractional OAM.  Further
to existing theories we have calculated the propagated fields in the paraxial and non-paraxial regime for an incident beam of finite width. The gradual formation of additional vortices in the region of low intensity is a phenomenon which is only visible in the non-paraxial solution. 

\section{Acknowledgements}

We are grateful to Michael Berry, Eric Yao, Jonathan Leach and Miles Padgett for helpful comments. This work was supported by the EPSRC under the grant GR S03898/01.

 %%
 %% Appendices
 %% 
 
 \appendix
 
 \section{Rotation of the orientation of the discontinuity}
 \label{app:unitaryrotation}
 
Quantum states with fractional OAM depend on $\alpha$, the orientation of the
phase discontinuity. To rotate the orientation we introduce a class of operators
$\hat{U}_m(\beta), m \in \mathbb{R}, \beta \in [0,2\pi)$ defined by their action on a state with fractional OAM:
\begin{equation}
\label{eq:rotopdef}
\hat{U}_m(\beta) | M' (\alpha) \rangle = \exp[\rmi(m-m')\beta] | M(\alpha \oplus \beta) \rangle,
\end{equation}
where $\beta$ is the angle through which the discontinuity is rotated.
The addition $\alpha \oplus \beta = (\alpha + \beta) \Mod 2 \pi$ yields a result in the range $[0,2\pi)$.
For $m=m'$, that is if $m$ is equal to the integer part of $M$, this operator is a pure rotation of the orientation $\alpha$, but for $m \neq m'$ the
rotated state acquires also a phase shift unless $\beta \Mod 2\pi = 0$. Consequently, the operator acts as identity operator if  $\beta$ is an integer multiple of $2\pi$. The eigenstates
of this operator are the integer OAM states. Acting with $\hat{U}_m(\beta)$ on 
an integer state $| m \rangle, m \in \mathbb{Z}$, results only in a phase shift:
\begin{equation}
\hat{U}_m(\beta) | m' \rangle = \exp[\rmi(m-m')\beta] | m' \rangle.
\end{equation}
All the eigenvalues have unit modulus and the operator 
$\hat{U}_m(\beta)$ is therefore unitary with $\hat{U}_m(\beta) = \hat{U}_m(-\beta) = \hat{U}^\dagger_m(\beta).$

From the definition of the operator in Eq. (\ref{eq:rotopdef}) it is obvious that the set of operators cannot
form a group under multiplication for arbitrary combinations of parameters $\beta$ and $m$. But for
a fixed $m$ the set of operators $\{ \hat{U}_m(\beta) \}_{\beta \in [0,2\pi)}$ does form a one-parameter group under multiplication with the group parameter $\beta$ \cite{gpI:sv:1990}. As proof 
we show that $\hat{U}_m(\beta') \hat{U}_m(\beta)$ can always be written as $\hat{U}_m(\gamma)$ with $\gamma \in [0,2\pi)$:
\begin{equation}
\begin{split}
\hat{U}_m(\beta') \hat{U}_m(\beta) | M' (\alpha) \rangle & = \hat{U}_m(\beta') \exp[\rmi (m-m')\beta] | M'(\alpha \oplus \beta) \rangle , \\
&= \exp[\rmi (m-m')(\beta + \beta')]| M'(\alpha \oplus \beta \oplus \beta') \rangle = \hat{U}_m(\beta \oplus \beta') | M'(\alpha) \rangle.
\end{split}
\end{equation}
Therefore $\hat{U}_m(\beta') \hat{U}_m(\beta) = \hat{U}_m(\beta \oplus \beta')$ as the modulo $2\pi$ addition $\beta \oplus \beta'$ is always in the interval $[0,2\pi)$. There is also a neutral element in form of the identity operator for $\beta = 0$. Moreover, to every operator $\hat{U}_m(\beta)$ exists an inverse element in the form 
$\hat{U}^\dagger_m(\beta) = \hat{U}_m(-\beta)$. 

The product of two operators with the same orientation but different values for the fractional orbital angular momentum can be combined to give
\begin{equation}
\hat{U}_m(\beta) \hat{U}_{m'}(\beta) = \exp[-\rmi (m+m')\beta] U_{m+m'}(\beta \oplus \beta).
\end{equation}
The set of operators $\{ \hat{U}_m(\beta)\}_{m \in \mathbb{Z}}$ does not form a  group under multiplication for $\beta \neq 0$. For $\beta = 0$ these operators are all identity operators.

We have introduced the operator $\hat{U}_m(\beta)$ to facilitate the calculation of the general overlap
$\langle M'(\alpha') | M(\alpha) \rangle$ in section \ref{sec:overlap}. The mathematical properties suggest, that the action of this operator on a state $| M'(\alpha) \rangle$ is independent of the fractional part $\mu'$. A physical implementation of this operator would be an optical device with two edge dislocations; one to compensate the original dislocation at an angle $\alpha$, the other to imprint the new phase discontinuity at an angle $\alpha \oplus \beta$. This implementation, however, would be specific for given fractional part $\mu'$. As the effect of the edge dislocation is completely specified by the fractional part $\mu'$ and the orientation $\alpha$ it would be possible to change the orientation of the state $|M'(\alpha)\rangle$ with an implementation of the operator $\hat{U}_m(\beta)$, where the integer part of $M'$  is different from $m$, which causes the overall phase factor in Eq. (\ref{eq:rotopdef}).

\end{document}